\documentclass[twocolumn,superscriptaddress,showpacs,pra,amsmath,amstex,amssymb,citeautoscript,longbibliography]{revtex4-1}
\pdfoutput=1
\usepackage{natbib}
\usepackage[english]{babel}
\usepackage{letltxmacro}
\usepackage{latexsym}
\LetLtxMacro{\ORIGselectlanguage}{\selectlanguage}
\makeatletter
\DeclareRobustCommand{\selectlanguage}[1]{%
  \@ifundefined{alias@\string#1}
    {\ORIGselectlanguage{#1}}
    {\begingroup\edef\x{\endgroup
       \noexpand\ORIGselectlanguage{\@nameuse{alias@#1}}}\x}%
}
\newcommand{\definelanguagealias}[2]{%
  \@namedef{alias@#1}{#2}%
}
\makeatother
\definelanguagealias{en}{english}
\definelanguagealias{English}{english}
\usepackage{graphicx}
\usepackage{amsmath}
\usepackage{amsfonts}
\usepackage{amssymb}
\usepackage{bm}
\usepackage{color}
\usepackage[percent]{overpic}
\usepackage{soul} 
\usepackage{amssymb}
\usepackage{wasysym}
\usepackage{dsfont}
\usepackage{float}

\usepackage{hyperref}
\hypersetup{
    bookmarks=false,         
    unicode=false,          
    pdftoolbar=false,        
    pdfmenubar=true,        
    pdffitwindow=false,     
    pdfstartview={FitH},    
    pdftitle={},    
    pdfauthor={Authors},     
    pdfsubject={},   
    pdfcreator={},   
    pdfproducer={}, 
    pdfkeywords={many-body localization} {matrix elements} {disordered systems}, 
    pdfnewwindow=true,      
    colorlinks=true,       
    linkcolor=black,          
    citecolor=blue,        
    filecolor=magenta,      
    urlcolor=blue           
}

\newcommand{\be}{\begin{equation}}
\newcommand{\ee}{\end{equation}}
\newcommand{\bea}{\begin{eqnarray}}
\newcommand{\eea}{\end{eqnarray}}

\newcommand{\beq}{\begin{eqnarray}}
\newcommand{\eeq}{\end{eqnarray}}

\usepackage{graphicx}
\usepackage[colorinlistoftodos]{todonotes}
\usepackage{verbatim}
\usepackage[normalem]{ulem}

\begin{document}

\title{
Ultrafast Variational Simulation of Non-trivial Quantum States with Long Range Interactions
}
\author{Wen Wei Ho}
\affiliation{Department of Physics, Harvard University, Cambridge, Massachusetts 02138, USA}

\author{Cheryne Jonay}
\affiliation{Perimeter Institute for Theoretical Physics, Waterloo, Ontario N2L 2Y5, Canada}

\author{Timothy H. Hsieh}
\affiliation{Perimeter Institute for Theoretical Physics, Waterloo, Ontario N2L 2Y5, Canada}

\date{\today}
\begin{abstract}
State preparation protocols ideally require as minimal operations as possible, in order to be implemented in near-term, potentially noisy quantum devices.
%
%
%
Motivated by long range interactions (LRIs) intrinsic to many present-day experimental platforms (trapped ions, Rydberg atom arrays, etc.),
we investigate the efficacy of   variationally simulating   non-trivial quantum states using the Variational Quantum-Classical Simulation (VQCS) protocol explored recently in [SciPost Phys.~6, 029 (2019)], in the presence of LRIs.
%
%
We show that this approach leads to extremely efficient state preparation: 
for example, Greene-Horne-Zeilinger (GHZ) states can be prepared with $O(1)$ iterations of the protocol, and a quantum critical point of the long range transverse field Ising model (TFIM) can be prepared with $>99\%$ fidelity on a 100 qubit system with {\it only one} iteration.
Furthermore, we show that VQCS with LRIs is a promising route for exploring generic points in the phase diagram of the long-range TFIM.
Our approach thus provides concrete, ultrafast protocols for quantum simulators equipped with long range interactions. 
\end{abstract}

\maketitle

\section{Introduction}

Rapid experimental progress in the   control of synthetic quantum systems, such as trapped ions \cite{Blatt12, Zhang2017_DPT, Islam583}, ultracold atoms \cite{Greiner2002, BlochColdAtoms, Bernien2017} and superconducting qubits \cite{Devoret13, Gambetta2017},
%
 {has ushered in the era of the so-called Noisy Intermediate-Scale Quantum (NISQ) technology \cite{Preskill2018}, where quantum devices of up to $50-100$ qubits can be coherently manipulated. This has} unlocked the potential for quantum computation \cite{Bloch15, Gambetta2017}, quantum sensing and metrology \cite{Leibfried1476, Giovannetti1330, RevModPhys.89.035002, 	ChoiMetrology}, and also the simulation of quantum many-body phases of matter \cite{Greiner2002,  PhysRevLett.111.185301, Aidelsburger2014, Islam2015, Bloch16, Smith2016, Choi16DTC, Zhang2017,  2018arXiv180610169C}.
%
Such tasks require the ability to create with good fidelities,  quantum states containing nontrivial entanglement, such as the Greene-Horne-Zeilinger (GHZ) state, quantum critical states, and topologically ordered states etc.
%
%
A central challenge is therefore  finding efficient state preparation protocols {that can be implemented in these noisy, imperfect quantum platforms}: ideally,   protocols should have as minimal a circuit depth as possible  to be realistically implemented, in order to suppress the errors that  accumulate during   runtime.
%

Recently, the Variational Quantum-Classical Simulation (VQCS) protocol was proposed as one such candidate \cite{2018arXiv180300026H}. 
%
  In short, the VQCS is a hybrid quantum-classical bang-bang protocol which specifically incorporates feedback, and  is motivated by the Quantum Approximate Optimization Algorithm (QAOA)  \cite{QAOA1, QAOA2} as well as various variational quantum eigensolvers \cite{variational,VQE}.  It works as follows: after initializing in an easily preparable state, a set of angles is fed into the quantum simulator, which specifies the durations for which time evolution between two different Hamiltonians is alternated between. Measurements are then performed to estimate the energy of the resulting state with respect to a target (generally quantum) Hamiltonian.  The energy cost function is subsequently optimized on a classical computer to yield a new set of angles, and the process is iterated until the cost function is minimized. 
With spatially local Hamiltonians and finite evolution times, the VQCS has been shown to be able  to transform trivial product states into GHZ, quantum critical, and topologically ordered states, with perfect fidelity and {iteration-}depths that scale as $O(N)$ where $N$ is the system's linear dimension \cite{2018arXiv180300026H}. 
Conceptually, the VQCS is an example of a ``shortcut to adiabaticity'', a direction in quantum state control that is actively being researched \cite{doi:10.1021/jp030708a, doi:10.1021/jp040647w, 1751-8121-42-36-365303, QuenchedPrep, Sels201619826}, as its operating principle is fundamentally different from     conventional adiabatic preparation schemes \cite{QAA1, QAA2, optimalControl1, optimalControl2, OptimalControl3, PhysRevX.7.021027}.

While  {an iteration}-depth scaling as $O(N)$ is efficient from a theoretical standpoint -- there exist fundamental speed limitations imposed by Lieb-Robinson bounds \cite{lieb1972, nielsen, PhysRevLett.97.050401, HastingsLR} constraining   unitary circuits utilizing spatially local Hamiltonians, it still presents challenges experimentally, especially in terms of scalability to a large number of qubits {in near-term devices}. 
This motivates the search for alternative ultrafast protocols.   
%
%
A  {possible} way to overcome these speed limitations is to utilize long-range interactions (LRIs)
 that are  naturally present in various experimental quantum simulator platforms, e.g. trapped ion systems (Coulomb interactions), Rydberg atom arrays (van der Waals interactions), etc. 
 \cite{Richerme2014, Zhang2017_DPT, Bernien2017}.    
%
With LRIs, entanglement and correlations can be built up between distant parts of the system in finite time \cite{LR_LR1, LR_LR2, LR_LR3},  potentially (though not obviously) allowing for a quick preparation of desired long-range correlated states.

To this end, in this work we explore how efficiently the VQCS protocol with long range interactions can prepare   non-trivial quantum states. 
%
Specifically, we consider in mind quantum simulators (digital or analog) that realize long-range  $\sim$\,$1/r^\alpha$ Ising interactions with tuneable range $\alpha$, motivated  {in large part} by trapped ion experimental setups.
%
We find that the VQCS protocol with LRIs can prepare GHZ and quantum critical states with $O(1)$ iterations.  In particular, in the limit of extremely long-range  interactions, the GHZ state can be prepared with only one (two) iteration(s) for odd (even) system sizes. Furthermore, the quantum critical point of the Lipkin-Meshkov-Glick model \cite{LMG} can be prepared with high fidelity very quickly (e.g. fidelity $>0.99$ for 100 spins after one iteration).  We also analyze how efficiently  the protocol can prepare points within the phase diagram of the long-range transverse field Ising model.  Our results thus demonstrate the utility of VQCS-protocols with LRIs for near-term, potentially noisy quantum simulators 
 to realize nontrivial many-body states of interest.     

\section{Variational Quantum-Classical Simulation (VQCS) protocol}  
 We quickly recapitulate the VQCS \cite{2018arXiv180300026H}. Our aim is to prepare a target state $|\psi_t\rangle$ with as high fidelity as possible, given resources available in a quantum simulator (either digital or analog) such as single qubit rotations and interactions between qubits, which we denote schematically by $H_1, H_2$.
 Usually, $|\psi_t\rangle$ will be taken to be the ground state of some target Hamiltonian $H_t$ which is a linear combination of $H_1, H_2$. Henceforth in this work,  we shall take $H_1 = -\sum_i X_i$, a global transverse field, but this can be relaxed. 

The VQCS starts off with an easily preparable initial state,
%
such as the unentangled ground state $|+\rangle$ of the paramagnet $H_1$. 
One then time evolves in an alternating fashion between $H_1$ and the ``interaction Hamiltonian'' $H_2$, for a total of $p$ iterations:
\begin{align}
| \psi (\vec{\gamma},\vec{\beta})  \rangle_p\! = \!e^{-i \beta_p H_1} e^{- i \gamma_p H_2} \cdots e^{-i \beta_1 H_1} e^{- i \gamma_1 H_2} |+ \rangle,
\label{eqn:QAOA}
\end{align}
with evolution times given by angles $(\vec{\gamma},\vec{\beta}) \equiv (\gamma_1,\cdots \gamma_p, \beta_1, \cdots, \beta_p)$. 
We label this protocol as VQCS$_p$.
%

As the goal is to closely approximate $H_t$'s ground state, one can seek to 
find the evolution times $(\vec{\gamma},\vec{\beta}) $ which minimize a given cost function $F_p(\vec{\gamma},\vec{\beta})$, usually taken to be the energy with respect to the target Hamiltonian $H_t$: 
\begin{align}
F_p(\vec{\gamma},\vec{\beta}) = {_p\langle} \psi (\vec{\gamma},\vec{\beta}) |H_t |\psi (\vec{\gamma},\vec{\beta}) \rangle_p.
\label{eqn:costFn}
\end{align}
Obviously, increasing $p$  can only improve the minimal value $F_p^*$, i.e.~$F_{p+1}^* \leq F_p^*$. 

In practice, such a protocol can be implemented in a hybrid setup involving a quantum simulator and a classical computer: one first feeds the quantum simulator an initial seed of angles,  producing a state $|\psi(\vec{\gamma},\vec{\beta})\rangle$. Then, leveraging upon single-site accesibility possible in many present-day quantum simulators, one measures correlations within the state and determines the cost function (\ref{eqn:costFn}), e.g.~the global energy. A classical computer is then used to obtain the next set of angles $(\vec{\gamma},\vec{\beta})$ to be fed into the quantum simulator, by means of an optimization algorithm such as gradient descent or a similar protocol. The entire process is then repeated until either the global minimum $F_p^*$ is found, or a desired energy/fidelity threshold is attained. 
As a matter of principle, the VQCS protocol is guaranteed to work in the limit of $p \to \infty$ for any finite size system (there always exists a finite gap), as an asymptotically slow adiabatic preparation scheme can always be trotterized to the form (\ref{eqn:QAOA}). 
However, non-trivial behavior and an improvement over adiabaticity can arise for small $p$, the regime of practical interest for experimental systems.
%
 
As an example, consider preparing the ground state of the one-dimensional nearest-neighbor transverse field Ising model (TFIM), a situation considered in \cite{2018arXiv180300026H}:
\begin{align}
H_{TFIM} = -\sum_{i=1}^N Z_i Z_{i+1} - h \sum_{i=1}^N X_i,
\label{eqn:HTFIM}
\end{align}
where $h$ parameterizes the field strength and $N$ is the number of qubits.  Given this $H_t$, a natural choice is $H_2 = -\sum_{i=1}^N Z_i Z_{i+1}$, which are interactions (approximately) naturally realizable in e.g.~trapped ions or Rydberg array simulators.
%
%
%
Indeed, in a previous work, it was shown that  such a VQCS$_{p^*}$ at $p^*=N/2$ can target with {\it perfect fidelity} the ground states of the model at $h=0,1$ (GHZ and quantum critical state, respectively) \cite{2018arXiv180300026H}. It was further conjectured and supported with numerical evidence that this result generalizes to all points $h \in \mathbb{R}$. 
%

\section{VQCS with long-range interactions (LRIs)}  
Despite impressive progress
 in the coherent control and manipulation of
   quantum systems today,  such platforms are inherently noisy, and so it 
is  desirable to have state preparation protocols that require as few iterations, and as short a runtime as possible.  
%
\textcolor{black}{ However, there fundamental exist speed limits (specifically, Lieb-Robinson bounds) in systems with local interactions to create a desired quantum state containing long-range entanglement -- the time taken is  $t \geq O(N)$ (as illustrated explicitly in the example above). Intuitively, this arises from the linear light cone $r \sim vt$ of information propagation that limits the speed at which spatially distant regions   entangle.}
%

Long-range interactions (LRIs) have less stringent speed limits \cite{LR_LR3} and can potentially dramatically speed up state preparation protocols.
%
%
We now show in the rest of the paper that the VQCS (\ref{eqn:QAOA}) with LRIs is a viable method for efficiently targeting nontrivial quantum states. 
%
%
%
We consider quantum simulators where  long-range Ising interactions 
$
H_2 = -\sum_{i<j}^N J_{ij} Z_i Z_j
$
 with $J_{ij} = \frac{J_0}{|i-j|^{\alpha}}$ for some power-law exponent $\alpha$, 
 can be realized, such as trapped ion setups or Rydberg atom array setups. Concretely, together with a readily applicable transverse field $H_1$,  we study the following prototypical, realizable effective Hamiltonians:
\begin{align}
H_t = -\sum_{i<j}^N J_{ij} Z_i Z_j  - {\cal N} h \sum_i^N X_i
\label{eqn:H}.
\end{align}
%
%
{In trapped ion setups, $\alpha$ can vary in principle between 0 and 3, with experiments having been conducted using $\alpha$ ranging from $0.67$ to $1.05$ \cite{Islam583}, while in Rydberg atom array setups, $\alpha = 6$.}   We have chosen $J_0 = 1$ and normalized (\ref{eqn:H}) in a standard way \cite{Zhang2017_DPT} so that ${\cal N} = \frac{1}{N-1} \sum_{i< j} J_{ij}$.  Note  that $\alpha \to \infty$ reduces to the nearest neighbor TFIM model with open boundary conditions.

\section{Ultrafast state preparation using VQCS with long-range interactions}

We now analyze  the small $\alpha$ regime of the Hamiltonian (\ref{eqn:H}) 
and show that VQCS protocols (\ref{eqn:QAOA}) using $H_1, H_2$ as defined above, with $p=O(1)$,  are sufficient to prepare certain target ground states.  
We  will restrict the VQCS parameter space to $\gamma_i \in [-\pi,\pi)$ and $\beta_i \in [0,\pi/2)$.  The former is motivated by experimental limitations on the evolution time, and the latter is because 
$e^{-i(\pi/2) H_X} \propto \prod_i X_i$ which is conserved throughout the evolution.

\begin{figure}[t]
\begin{center}
\includegraphics[width=0.45\textwidth]{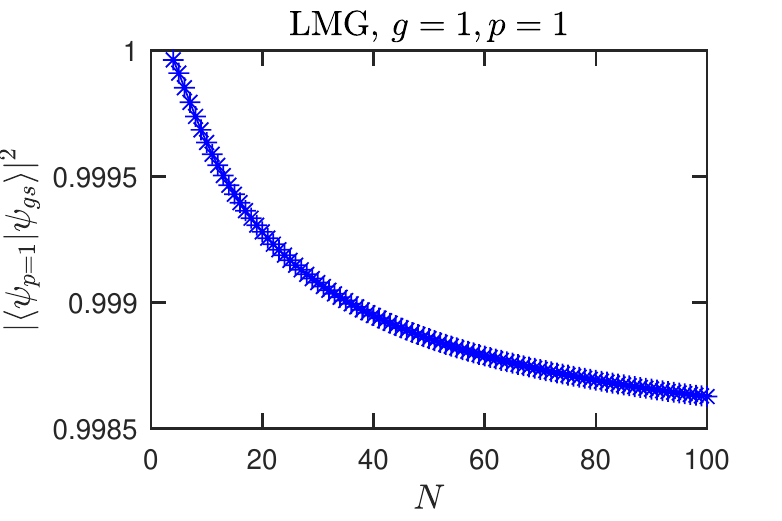}
\caption{Fidelity $|\langle \psi_{QCP}|\psi(\vec{\gamma},\vec{\beta})\rangle_{p=1}|^2$ in preparation of LMG critical state as a function of system size $N$, for VQCS depth $p=1$. 
Remarkably, even at $N= 100$, the fidelity is very close to unity $(>99\%)$.
}
\label{fig:FidelityLMGcrit}
\end{center}
\end{figure}


\subsection{ GHZ state preparation, $\alpha = 0$}
First consider the case $\alpha=0$ of (\ref{eqn:H}), in which the $N$ qubits interact in an all-to-all fashion.  Then, up to an overall multiplicative factor and also an inconsequential shift in energy, (\ref{eqn:H}) is equivalent to the Lipkin-Meshkov-Glick (LMG) model
\begin{align}
H_{LMG} = -\frac{2}{N} S_z^2 - 2g S_x,
\label{eqn:HLMG}
\end{align}
where the total spin operators are $S_z = \sum_i^N Z_i/2$ and $S_x = \sum_i^N X_i/2$, and $g=h/2$.  As is the case with the nearest-neighbor TFIM model, its ground states are ferromagnetic GHZ states at $g=0$, and a quantum phase transition at $g=1$ separates the ferromagnet from the paramagnetic phase.

We claim a $p$\,$=$\,$O(1)$ VQCS circuit suffices to produce the ground state of (\ref{eqn:HLMG}) at $g$\,$=$\,$0$, i.e.~the GHZ state. To see this, we explicitly derive the energy cost function (\ref{eqn:costFn}) for the LMG model with VQCS$_{p=1}$:
\begin{align}
\label{eqn:p1CostFn}
& F_{p=1}(\gamma, \beta) = -\frac{N-1}{4} \Big(\sin(2\beta)^2 (1-\cos(4\gamma)^{N-2})  +  \\ \nonumber
& 2\sin(4\beta)  \sin(2\gamma)\cos(2\gamma)^{N-2}\Big) 
-gN\cos(2\gamma)^{N-1}-1/2
\end{align}
 (see Appendix A for the derivation). From the above, it is evident that for odd $N$, the ground state energy of $H_{LMG}|_{g=0}$, namely $E_0$\,$=$\,$-N/2$, can be achieved with angles $(\gamma, \beta)= (\pi/4, \pi/4)$.  
In other words, the ferromagnetic GHZ state, a state with macroscopic superposition of entanglement $(1/\sqrt{2})(|0...0\rangle + |1...1\rangle)$, can be created with just two operations:
\begin{align}
|GHZ\rangle = e^{-i (\pi/4) H_X} e^{- i (\pi/4) H_I} |+ \rangle.
\end{align}

We note that there exist various existing preparation schemes that create macroscopic GHZ states, one of which is the Molmer-Sorenson(MS) protocol involving time evolution with $S_x^2$ \cite{PhysRevA.62.022311, PhysRevLett.106.130506}.
Although the VQCS protocol discussed above somewhat resembles the MS protocol, there are several differences: 
MS begins with the (Ising symmetry broken) ground state $|0\cdots0\rangle$ and for odd system sizes, involves time evolution with $S_x^2$ and $S_x$, to produce a GHZ state with a relative phase between the cat states. A single qubit gate, or alternatively time evolution with $S_z$, can remove the relative phase.

%

The distinction between MS and our protocol is most manifest for even system sizes, in which we find that the GHZ state is instead achieved with perfect fidelity with a $p$\,$=$\,$2$ VQCS protocol: 
\begin{align} 
\label{eqqq}
|GHZ\rangle =e^{-i (\pi/4) H_X} e^{- i (\pi/8) H_I}  e^{-i (3\pi/4N) H_X} e^{- i (\pi/4) H_I} |+ \rangle. \nonumber
\end{align}
Note that from (\ref{eqn:p1CostFn}), there is no range of parameters that give perfect fidelity for $p$\,$=$\,$1$ for even $N$.
We show in Appendix B the derivation of the above result.  
These results already demonstrate the utility of VQCS with LRIs: they enable ultrafast preparation of a macroscopic GHZ state, with perfect fidelity.

\begin{figure}[t]
\begin{center}
\includegraphics[width=0.5\textwidth]{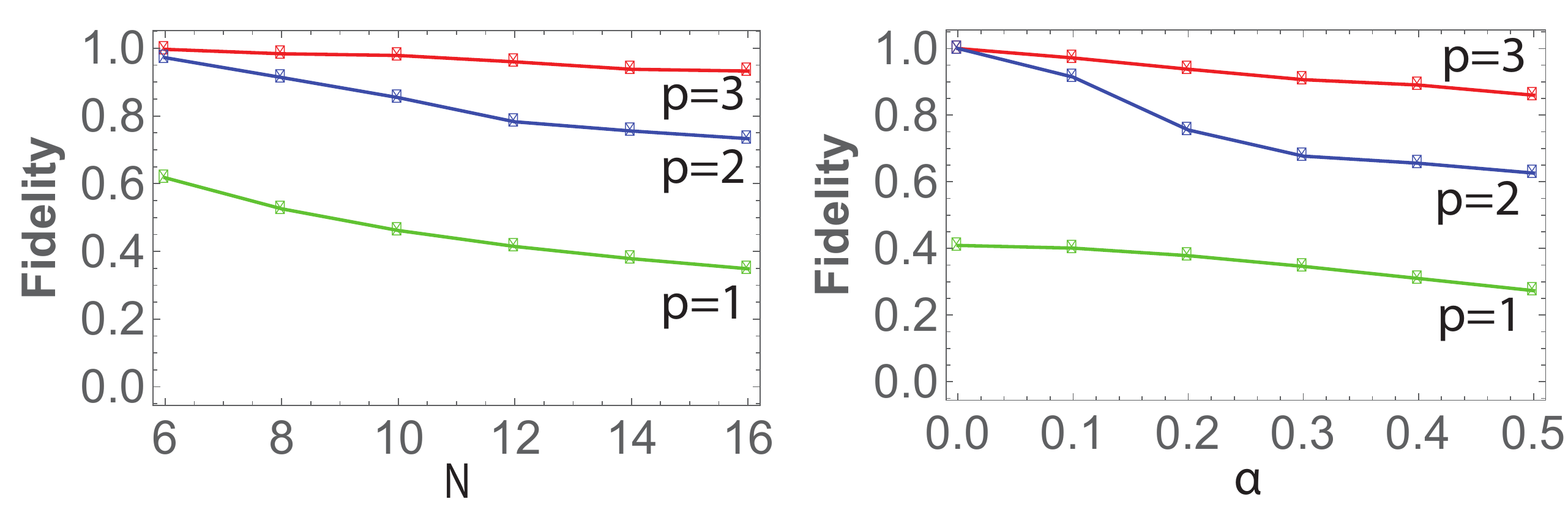}
\caption{(Left) Fidelity in preparation of GHZ state as a function of system size $N$, for $\alpha$\,$=$\,$0.2$.
The fidelity decreases with increasing $N$; however, this can be compensated by going to higher $p$s. 
 (Right) Fidelity in preparation of GHZ state as a function of $\alpha$, for $N$\,$=$\,$14$. 
}
\label{fig:Print}
\end{center}
\end{figure}

\begin{figure*}[t]
\includegraphics[width=1.02\textwidth]{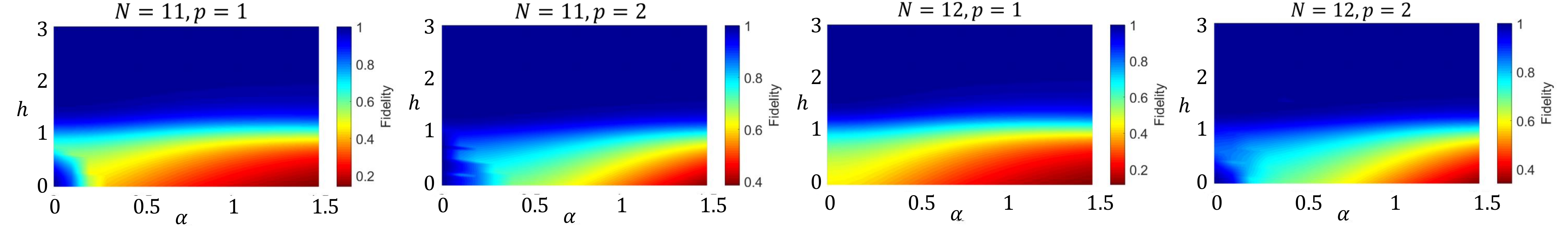}
\caption{Heat map depicting fidelities in the state preparation of the ground state of the long-range TFIM  Eq.~(\ref{eqn:HTFIM})  for system sizes  $N = 11$, and for 12, $p = 1,2$. Blue (red) regions demarcate regions of high (low) fidelity. As noted before, there is an odd-even system size distinction near $\alpha \approx 0$, $h \approx 0$ for $p = 1$, but this distinction vanishes at the next level, $p = 2$.  
}
\label{fig:Fig 3}
\end{figure*}

 \subsection{ Quantum critical state preparation, $\alpha = 0$}

Besides the GHZ state, we find the approach 
can target many other interesting states. 
One state of particular interest is the quantum critical point of the LMG model at $g=1$, a highly correlated state $|\psi_{QCP}\rangle$.  By numerically optimizing for the energy of (\ref{eqn:HLMG}) at $g = 1$, we find that the $p$\,$=$\,$1$ VQCS protocol  is already sufficient to achieve the critical state with extremely high fidelity $|\langle \psi_{QCP}|\psi(\vec{\gamma},\vec{\beta})\rangle_{p=1}|^2$, even for very large system sizes ($> 99 \%$ at $\sim$100 qubits) (see Fig.~\ref{fig:FidelityLMGcrit}).  This is a remarkably efficient protocol for preparing a critical state.

Note that the gap at the critical point of the LMG model scales as $\Delta \propto N^{-1/3}$ \cite{LMG}, and thus the adiabatic algorithm requires $O(N^{1/3})$ time to prepare the GHZ and critical states.  However, in order to make a \textcolor{black}{meaningful} comparison of the VQCS with the adiabatic algorithm, we would need to scale down the exchange interaction in (5) (i.e.~$J_0$) by $N$, and thus the total preparation time for the VQCS protocol in this convention would go as $N$.  Our intention is not to make this theoretical comparison, but instead to make contact with existing experiments.  In the trapped ion setups of \cite{Zhang2017_DPT}, the nearest-neighbor exchange interaction $J_0$ does not necessarily decrease with system size; for example, it is $(0.82, 0.56, 0.38, 0.65) $kHz for $N$\,$=$\,$(8$\,$,$\,$12$\,$,$\,$16$\,$,$\,$53)$ respectively.   Hence, the interactions in (5) are reasonable in near-term trapped ion experiments and lead to $O(1)$ preparation time for the GHZ and critical states.  The simplicity and discreteness of the protocols we have presented may offer advantages over the adiabatic algorithm with or without counter-diabatic terms \cite{PhysRevLett.111.100502, PhysRevLett.114.177206, Sels201619826, PhysRevLett.118.100601}.

\subsection{ GHZ preparation with finite $\alpha$}
 In practice, there may be challenges in realizing strictly all-to-all ($\alpha$\,$=$\,$0$) Ising interactions, and therefore we analyze how well a  finite $\alpha$ VQCS protocol can prepare the GHZ state.

In the left panel of Fig.~2, we fix $\alpha$\,$=$\,$0.2$ and show the fidelity with GHZ state achieved for VQCS$_{p=1,2,3}$ for even system sizes.
 As expected, VQCS$_{p=2}$ no longer prepares the state with perfect fidelity unlike the $\alpha = 0$ case, but this can be addressed with further iterations of VQCS. In particular, note the high fidelities achieved by $p$\,$=$\,$3$ for system sizes up to $N = 16$.
As LRIs establish correlations between spins separated by a distance $r$ in time $O(r^\alpha)$, which surpasses the light cone bound for local interactions \cite{LR_LR1, LR_LR2, LR_LR3}, we expect that the depth required to prepare the state with some fixed error in fidelity scales as $O(N^{\alpha'}), \alpha'<1$.
 Also plotted is the optimal fidelity for different values of $\alpha$ and for fixed system size $N$\,$=$\,$14$; the results indicate that longer-range interactions (smaller $\alpha$) tend to be more effective in targeting the desired state, and that
%
 errors can be effectively reduced using further VQCS iterations \footnote{We note that the numerical optimization in our simulations may output local minima in the cost function, and thus the results presented in Fig. 2 are lower bounds on the optimal fidelities.}.

\subsection{Phase diagram of the long-range TFIM}
  Finally, we explore how well the VQCS protocol with LRIs can prepare the ground states at generic points in the phase diagram of the long-range TFIM model (\ref{eqn:H}). 
It is known that the long-range TFIM supports a ferromagnetic-paramagnetic ground state quantum phase transition for any value of $\alpha$, upon tuning $h$. In the limit $\alpha$\,$\to$\,$0$, the critical  field $h_c$\,$=$\,$2$, while in the limit $\alpha$\,$\to$\,$\infty$, the critical  field $h_c$\,$=$\,$1$. For intermediate values of $\alpha$, previous works have attempted to map out how $h_c$ varies (see e.g. \cite{Carr, schmidt} and \cite{PhysRevLett.109.267203, lrisingplus} for the antiferromagnetic model). 

 Plotted in Fig. 3 are the fidelities obtained from VQCS as a function of transverse field $h$ and interaction range $\alpha$, for $N=11,12$ and for $p=1,2$.
  As expected,  at the system sizes considered, the VQCS with LRIs is able to target ground states within the paramagnetic phase (large $h$) relatively easily, while for the ferromagnetic phase (small $h$) it becomes more difficult to prepare, especially as the interactions become more short-ranged (larger $\alpha$).  
 Note that for large $\alpha$,  the region where state preparation is difficult (red region) is separated from the region   where state preparation is easy (blue region) by $h \approx 1$, which agrees with the critical point $h_c$ of the nearest-neighbor TFIM, which is realized in the asymptotic limit $\alpha\rightarrow \infty$.


As the small $\alpha$ regime of this model is somewhat challenging for numerical studies \cite{Carr}, it serves as a venue in which the quantum-classical hybrid implementation of VQCS could provide valuable input. Moreover, the small $\alpha$ window is precisely the regime in which the VQCS approach requires only a few iterations.


\section{Discussion and conclusion} 
We have shown that VQCS-type protocols with long-range interactions allow for ultra-fast state preparation.  In particular, the Ising-symmetric GHZ state can be prepared exactly at finite depth $p=1 (2)$ for odd (even) system sizes.   We have also demonstrated that other states of interest, for example the quantum critical point of the Lipkin-Meshkov-Glick model, can also be prepared very efficiently.  
{
More broadly, since the VQCS protocol is very general and not only restricted to the states considered (see Eq.~(\ref{eqn:QAOA})), our results suggest that VQCS with LRIs is a promising and viable state preparation protocol that can be utilized to target other nontrivial states of interest, potentially also allowing for their efficient preparation.
}

VQCS with long range interactions thus provides an opportunity for near-term simulators to prepare non-trivial states with very high fidelity, and to shed light on areas of phase diagrams that are challenging for numerics.  The simplicity and efficiency of the protocols make them particularly well-suited for near-term quantum devices endowed with long-range interactions, such as trapped ion or Rydberg atom arrays. 
\\

{\it Note added:} While completing this manuscript, we became aware of related, variational state preparation works \cite{zoller18} and \cite{bapat}.

\begin{acknowledgments}
{We thank R. Islam, C. Monroe, and B. Yoshida for useful discussions.  WWH is supported by the Gordon and Betty Moore Foundation’s EPiQS Initiative through Grant No.~GBMF4306.  Research at Perimeter Institute
is supported by the Government of Canada through Industry
Canada and by the Province of Ontario through
the Ministry of Research and Innovation.  This work was performed in part at the Aspen Center for Physics, which is supported by National Science Foundation grant PHY-1607611.}
\end{acknowledgments}



\bibliography{refs}

\begin{thebibliography}{59}%
\makeatletter
\providecommand \@ifxundefined [1]{%
 \@ifx{#1\undefined}
}%
\providecommand \@ifnum [1]{%
 \ifnum #1\expandafter \@firstoftwo
 \else \expandafter \@secondoftwo
 \fi
}%
\providecommand \@ifx [1]{%
 \ifx #1\expandafter \@firstoftwo
 \else \expandafter \@secondoftwo
 \fi
}%
\providecommand \natexlab [1]{#1}%
\providecommand \enquote  [1]{``#1''}%
\providecommand \bibnamefont  [1]{#1}%
\providecommand \bibfnamefont [1]{#1}%
\providecommand \citenamefont [1]{#1}%
\providecommand \href@noop [0]{\@secondoftwo}%
\providecommand \href [0]{\begingroup \@sanitize@url \@href}%
\providecommand \@href[1]{\@@startlink{#1}\@@href}%
\providecommand \@@href[1]{\endgroup#1\@@endlink}%
\providecommand \@sanitize@url [0]{\catcode `\\12\catcode `\$12\catcode
  `\&12\catcode `\#12\catcode `\^12\catcode `\_12\catcode `\%12\relax}%
\providecommand \@@startlink[1]{}%
\providecommand \@@endlink[0]{}%
\providecommand \url  [0]{\begingroup\@sanitize@url \@url }%
\providecommand \@url [1]{\endgroup\@href {#1}{\urlprefix }}%
\providecommand \urlprefix  [0]{URL }%
\providecommand \Eprint [0]{\href }%
\providecommand \doibase [0]{http://dx.doi.org/}%
\providecommand \selectlanguage [0]{\@gobble}%
\providecommand \bibinfo  [0]{\@secondoftwo}%
\providecommand \bibfield  [0]{\@secondoftwo}%
\providecommand \translation [1]{[#1]}%
\providecommand \BibitemOpen [0]{}%
\providecommand \bibitemStop [0]{}%
\providecommand \bibitemNoStop [0]{.\EOS\space}%
\providecommand \EOS [0]{\spacefactor3000\relax}%
\providecommand \BibitemShut  [1]{\csname bibitem#1\endcsname}%
\let\auto@bib@innerbib\@empty
\bibitem [{\citenamefont {Blatt}\ and\ \citenamefont {Roos}(2012)}]{Blatt12}%
  \BibitemOpen
  \bibfield  {author} {\bibinfo {author} {\bibfnamefont {R.}~\bibnamefont
  {Blatt}}\ and\ \bibinfo {author} {\bibfnamefont {C.~F.}\ \bibnamefont
  {Roos}},\ }\bibfield  {title} {\enquote {\bibinfo {title} {Quantum
  simulations with trapped ions},}\ }\href@noop {} {\bibfield  {journal}
  {\bibinfo  {journal} {Nature Physics}\ }\textbf {\bibinfo {volume} {8}},\
  \bibinfo {pages} {277--284} (\bibinfo {year} {2012})}\BibitemShut {NoStop}%
\bibitem [{\citenamefont {Zhang}\ \emph
  {et~al.}(2017{\natexlab{a}})\citenamefont {Zhang}, \citenamefont {Pagano},
  \citenamefont {Hess}, \citenamefont {Kyprianidis}, \citenamefont {Becker},
  \citenamefont {Kaplan}, \citenamefont {Gorshkov}, \citenamefont {Gong},\ and\
  \citenamefont {Monroe}}]{Zhang2017_DPT}%
  \BibitemOpen
  \bibfield  {author} {\bibinfo {author} {\bibfnamefont {J.}~\bibnamefont
  {Zhang}}, \bibinfo {author} {\bibfnamefont {G.}~\bibnamefont {Pagano}},
  \bibinfo {author} {\bibfnamefont {P.~W.}\ \bibnamefont {Hess}}, \bibinfo
  {author} {\bibfnamefont {A.}~\bibnamefont {Kyprianidis}}, \bibinfo {author}
  {\bibfnamefont {P.}~\bibnamefont {Becker}}, \bibinfo {author} {\bibfnamefont
  {H.}~\bibnamefont {Kaplan}}, \bibinfo {author} {\bibfnamefont {A.~V.}\
  \bibnamefont {Gorshkov}}, \bibinfo {author} {\bibfnamefont {Z.-X.}\
  \bibnamefont {Gong}}, \ and\ \bibinfo {author} {\bibfnamefont
  {C.}~\bibnamefont {Monroe}},\ }\bibfield  {title} {\enquote {\bibinfo {title}
  {Observation of a many-body dynamical phase transition with a 53-qubit
  quantum simulator},}\ }\href {http://dx.doi.org/10.1038/nature24654}
  {\bibfield  {journal} {\bibinfo  {journal} {Nature}\ }\textbf {\bibinfo
  {volume} {551}},\ \bibinfo {pages} {601 EP --} (\bibinfo {year}
  {2017}{\natexlab{a}})}\BibitemShut {NoStop}%
\bibitem [{\citenamefont {Islam}\ \emph {et~al.}(2013)\citenamefont {Islam},
  \citenamefont {Senko}, \citenamefont {Campbell}, \citenamefont {Korenblit},
  \citenamefont {Smith}, \citenamefont {Lee}, \citenamefont {Edwards},
  \citenamefont {Wang}, \citenamefont {Freericks},\ and\ \citenamefont
  {Monroe}}]{Islam583}%
  \BibitemOpen
  \bibfield  {author} {\bibinfo {author} {\bibfnamefont {R.}~\bibnamefont
  {Islam}}, \bibinfo {author} {\bibfnamefont {C.}~\bibnamefont {Senko}},
  \bibinfo {author} {\bibfnamefont {W.~C.}\ \bibnamefont {Campbell}}, \bibinfo
  {author} {\bibfnamefont {S.}~\bibnamefont {Korenblit}}, \bibinfo {author}
  {\bibfnamefont {J.}~\bibnamefont {Smith}}, \bibinfo {author} {\bibfnamefont
  {A.}~\bibnamefont {Lee}}, \bibinfo {author} {\bibfnamefont {E.~E.}\
  \bibnamefont {Edwards}}, \bibinfo {author} {\bibfnamefont {C.-C.~J.}\
  \bibnamefont {Wang}}, \bibinfo {author} {\bibfnamefont {J.~K.}\ \bibnamefont
  {Freericks}}, \ and\ \bibinfo {author} {\bibfnamefont {C.}~\bibnamefont
  {Monroe}},\ }\bibfield  {title} {\enquote {\bibinfo {title} {Emergence and
  frustration of magnetism with variable-range interactions in a quantum
  simulator},}\ }\href {\doibase 10.1126/science.1232296} {\bibfield  {journal}
  {\bibinfo  {journal} {Science}\ }\textbf {\bibinfo {volume} {340}},\ \bibinfo
  {pages} {583--587} (\bibinfo {year} {2013})}\BibitemShut {NoStop}%
\bibitem [{\citenamefont {Greiner}\ \emph {et~al.}(2002)\citenamefont
  {Greiner}, \citenamefont {Mandel}, \citenamefont {Esslinger}, \citenamefont
  {H{\"a}nsch},\ and\ \citenamefont {Bloch}}]{Greiner2002}%
  \BibitemOpen
  \bibfield  {author} {\bibinfo {author} {\bibfnamefont {Markus}\ \bibnamefont
  {Greiner}}, \bibinfo {author} {\bibfnamefont {Olaf}\ \bibnamefont {Mandel}},
  \bibinfo {author} {\bibfnamefont {Tilman}\ \bibnamefont {Esslinger}},
  \bibinfo {author} {\bibfnamefont {Theodor~W.}\ \bibnamefont {H{\"a}nsch}}, \
  and\ \bibinfo {author} {\bibfnamefont {Immanuel}\ \bibnamefont {Bloch}},\
  }\bibfield  {title} {\enquote {\bibinfo {title} {Quantum phase transition
  from a superfluid to a mott insulator in a gas of ultracold atoms},}\ }\href
  {http://dx.doi.org/10.1038/415039a} {\bibfield  {journal} {\bibinfo
  {journal} {Nature}\ }\textbf {\bibinfo {volume} {415}},\ \bibinfo {pages} {39
  EP --} (\bibinfo {year} {2002})},\ \bibinfo {note} {article}\BibitemShut
  {NoStop}%
\bibitem [{\citenamefont {Bloch}\ \emph {et~al.}(2008)\citenamefont {Bloch},
  \citenamefont {Dalibard},\ and\ \citenamefont {Zwerger}}]{BlochColdAtoms}%
  \BibitemOpen
  \bibfield  {author} {\bibinfo {author} {\bibfnamefont {Immanuel}\
  \bibnamefont {Bloch}}, \bibinfo {author} {\bibfnamefont {Jean}\ \bibnamefont
  {Dalibard}}, \ and\ \bibinfo {author} {\bibfnamefont {Wilhelm}\ \bibnamefont
  {Zwerger}},\ }\bibfield  {title} {\enquote {\bibinfo {title} {Many-body
  physics with ultracold gases},}\ }\href {\doibase 10.1103/RevModPhys.80.885}
  {\bibfield  {journal} {\bibinfo  {journal} {Rev. Mod. Phys.}\ }\textbf
  {\bibinfo {volume} {80}},\ \bibinfo {pages} {885--964} (\bibinfo {year}
  {2008})}\BibitemShut {NoStop}%
\bibitem [{\citenamefont {Bernien}\ \emph {et~al.}(2017)\citenamefont
  {Bernien}, \citenamefont {Schwartz}, \citenamefont {Keesling}, \citenamefont
  {Levine}, \citenamefont {Omran}, \citenamefont {Pichler}, \citenamefont
  {Choi}, \citenamefont {Zibrov}, \citenamefont {Endres}, \citenamefont
  {Greiner}, \citenamefont {Vuletic},\ and\ \citenamefont
  {Lukin}}]{Bernien2017}%
  \BibitemOpen
  \bibfield  {author} {\bibinfo {author} {\bibfnamefont {Hannes}\ \bibnamefont
  {Bernien}}, \bibinfo {author} {\bibfnamefont {Sylvain}\ \bibnamefont
  {Schwartz}}, \bibinfo {author} {\bibfnamefont {Alexander}\ \bibnamefont
  {Keesling}}, \bibinfo {author} {\bibfnamefont {Harry}\ \bibnamefont
  {Levine}}, \bibinfo {author} {\bibfnamefont {Ahmed}\ \bibnamefont {Omran}},
  \bibinfo {author} {\bibfnamefont {Hannes}\ \bibnamefont {Pichler}}, \bibinfo
  {author} {\bibfnamefont {Soonwon}\ \bibnamefont {Choi}}, \bibinfo {author}
  {\bibfnamefont {Alexander~S.}\ \bibnamefont {Zibrov}}, \bibinfo {author}
  {\bibfnamefont {Manuel}\ \bibnamefont {Endres}}, \bibinfo {author}
  {\bibfnamefont {Markus}\ \bibnamefont {Greiner}}, \bibinfo {author}
  {\bibfnamefont {Vladan}\ \bibnamefont {Vuletic}}, \ and\ \bibinfo {author}
  {\bibfnamefont {Mikhail~D.}\ \bibnamefont {Lukin}},\ }\bibfield  {title}
  {\enquote {\bibinfo {title} {Probing many-body dynamics on a 51-atom quantum
  simulator},}\ }\href {http://dx.doi.org/10.1038/nature24622} {\bibfield
  {journal} {\bibinfo  {journal} {Nature}\ }\textbf {\bibinfo {volume} {551}},\
  \bibinfo {pages} {579 EP --} (\bibinfo {year} {2017})},\ \bibinfo {note}
  {article}\BibitemShut {NoStop}%
\bibitem [{\citenamefont {Devoret}\ and\ \citenamefont
  {Schoelkopf}(2013)}]{Devoret13}%
  \BibitemOpen
  \bibfield  {author} {\bibinfo {author} {\bibfnamefont {M.~H.}\ \bibnamefont
  {Devoret}}\ and\ \bibinfo {author} {\bibfnamefont {R.~J.}\ \bibnamefont
  {Schoelkopf}},\ }\bibfield  {title} {\enquote {\bibinfo {title}
  {Superconducting circuits for quantum information: an outlook},}\ }\href@noop
  {} {\bibfield  {journal} {\bibinfo  {journal} {Science}\ }\textbf {\bibinfo
  {volume} {339}},\ \bibinfo {pages} {1169} (\bibinfo {year}
  {2013})}\BibitemShut {NoStop}%
\bibitem [{\citenamefont {Gambetta}\ \emph {et~al.}(2017)\citenamefont
  {Gambetta}, \citenamefont {Chow},\ and\ \citenamefont
  {Steffen}}]{Gambetta2017}%
  \BibitemOpen
  \bibfield  {author} {\bibinfo {author} {\bibfnamefont {Jay~M.}\ \bibnamefont
  {Gambetta}}, \bibinfo {author} {\bibfnamefont {Jerry~M.}\ \bibnamefont
  {Chow}}, \ and\ \bibinfo {author} {\bibfnamefont {Matthias}\ \bibnamefont
  {Steffen}},\ }\bibfield  {title} {\enquote {\bibinfo {title} {Building
  logical qubits in a superconducting quantum computing system},}\ }\href
  {\doibase 10.1038/s41534-016-0004-0} {\bibfield  {journal} {\bibinfo
  {journal} {npj Quantum Information}\ }\textbf {\bibinfo {volume} {3}},\
  \bibinfo {pages} {2} (\bibinfo {year} {2017})}\BibitemShut {NoStop}%
\bibitem [{\citenamefont {Preskill}(2018)}]{Preskill2018}%
  \BibitemOpen
  \bibfield  {author} {\bibinfo {author} {\bibfnamefont {John}\ \bibnamefont
  {Preskill}},\ }\bibfield  {title} {\enquote {\bibinfo {title} {Quantum
  {C}omputing in the {NISQ} era and beyond},}\ }\href {\doibase
  10.22331/q-2018-08-06-79} {\bibfield  {journal} {\bibinfo  {journal}
  {{Quantum}}\ }\textbf {\bibinfo {volume} {2}},\ \bibinfo {pages} {79}
  (\bibinfo {year} {2018})}\BibitemShut {NoStop}%
\bibitem [{\citenamefont {Schreiber}\ \emph {et~al.}(2015)\citenamefont
  {Schreiber}, \citenamefont {Hodgman}, \citenamefont {Bordia}, \citenamefont
  {L{\"u}schen}, \citenamefont {Fischer}, \citenamefont {Vosk}, \citenamefont
  {Altman}, \citenamefont {Schneider},\ and\ \citenamefont {Bloch}}]{Bloch15}%
  \BibitemOpen
  \bibfield  {author} {\bibinfo {author} {\bibfnamefont {Michael}\ \bibnamefont
  {Schreiber}}, \bibinfo {author} {\bibfnamefont {Sean~S.}\ \bibnamefont
  {Hodgman}}, \bibinfo {author} {\bibfnamefont {Pranjal}\ \bibnamefont
  {Bordia}}, \bibinfo {author} {\bibfnamefont {Henrik~P.}\ \bibnamefont
  {L{\"u}schen}}, \bibinfo {author} {\bibfnamefont {Mark~H.}\ \bibnamefont
  {Fischer}}, \bibinfo {author} {\bibfnamefont {Ronen}\ \bibnamefont {Vosk}},
  \bibinfo {author} {\bibfnamefont {Ehud}\ \bibnamefont {Altman}}, \bibinfo
  {author} {\bibfnamefont {Ulrich}\ \bibnamefont {Schneider}}, \ and\ \bibinfo
  {author} {\bibfnamefont {Immanuel}\ \bibnamefont {Bloch}},\ }\bibfield
  {title} {\enquote {\bibinfo {title} {Observation of many-body localization of
  interacting fermions in a quasirandom optical lattice},}\ }\href {\doibase
  10.1126/science.aaa7432} {\bibfield  {journal} {\bibinfo  {journal}
  {Science}\ }\textbf {\bibinfo {volume} {349}},\ \bibinfo {pages} {842--845}
  (\bibinfo {year} {2015})}\BibitemShut {NoStop}%
\bibitem [{\citenamefont {Leibfried}\ \emph {et~al.}(2004)\citenamefont
  {Leibfried}, \citenamefont {Barrett}, \citenamefont {Schaetz}, \citenamefont
  {Britton}, \citenamefont {Chiaverini}, \citenamefont {Itano}, \citenamefont
  {Jost}, \citenamefont {Langer},\ and\ \citenamefont
  {Wineland}}]{Leibfried1476}%
  \BibitemOpen
  \bibfield  {author} {\bibinfo {author} {\bibfnamefont {D.}~\bibnamefont
  {Leibfried}}, \bibinfo {author} {\bibfnamefont {M.~D.}\ \bibnamefont
  {Barrett}}, \bibinfo {author} {\bibfnamefont {T.}~\bibnamefont {Schaetz}},
  \bibinfo {author} {\bibfnamefont {J.}~\bibnamefont {Britton}}, \bibinfo
  {author} {\bibfnamefont {J.}~\bibnamefont {Chiaverini}}, \bibinfo {author}
  {\bibfnamefont {W.~M.}\ \bibnamefont {Itano}}, \bibinfo {author}
  {\bibfnamefont {J.~D.}\ \bibnamefont {Jost}}, \bibinfo {author}
  {\bibfnamefont {C.}~\bibnamefont {Langer}}, \ and\ \bibinfo {author}
  {\bibfnamefont {D.~J.}\ \bibnamefont {Wineland}},\ }\bibfield  {title}
  {\enquote {\bibinfo {title} {Toward heisenberg-limited spectroscopy with
  multiparticle entangled states},}\ }\href {\doibase 10.1126/science.1097576}
  {\bibfield  {journal} {\bibinfo  {journal} {Science}\ }\textbf {\bibinfo
  {volume} {304}},\ \bibinfo {pages} {1476--1478} (\bibinfo {year}
  {2004})}\BibitemShut {NoStop}%
\bibitem [{\citenamefont {Giovannetti}\ \emph {et~al.}(2004)\citenamefont
  {Giovannetti}, \citenamefont {Lloyd},\ and\ \citenamefont
  {Maccone}}]{Giovannetti1330}%
  \BibitemOpen
  \bibfield  {author} {\bibinfo {author} {\bibfnamefont {Vittorio}\
  \bibnamefont {Giovannetti}}, \bibinfo {author} {\bibfnamefont {Seth}\
  \bibnamefont {Lloyd}}, \ and\ \bibinfo {author} {\bibfnamefont {Lorenzo}\
  \bibnamefont {Maccone}},\ }\bibfield  {title} {\enquote {\bibinfo {title}
  {Quantum-enhanced measurements: Beating the standard quantum limit},}\ }\href
  {\doibase 10.1126/science.1104149} {\bibfield  {journal} {\bibinfo  {journal}
  {Science}\ }\textbf {\bibinfo {volume} {306}},\ \bibinfo {pages} {1330--1336}
  (\bibinfo {year} {2004})}\BibitemShut {NoStop}%
\bibitem [{\citenamefont {Degen}\ \emph {et~al.}(2017)\citenamefont {Degen},
  \citenamefont {Reinhard},\ and\ \citenamefont
  {Cappellaro}}]{RevModPhys.89.035002}%
  \BibitemOpen
  \bibfield  {author} {\bibinfo {author} {\bibfnamefont {C.~L.}\ \bibnamefont
  {Degen}}, \bibinfo {author} {\bibfnamefont {F.}~\bibnamefont {Reinhard}}, \
  and\ \bibinfo {author} {\bibfnamefont {P.}~\bibnamefont {Cappellaro}},\
  }\bibfield  {title} {\enquote {\bibinfo {title} {Quantum sensing},}\ }\href
  {\doibase 10.1103/RevModPhys.89.035002} {\bibfield  {journal} {\bibinfo
  {journal} {Rev. Mod. Phys.}\ }\textbf {\bibinfo {volume} {89}},\ \bibinfo
  {pages} {035002} (\bibinfo {year} {2017})}\BibitemShut {NoStop}%
\bibitem [{\citenamefont {{Choi}}\ \emph
  {et~al.}(2018{\natexlab{a}})\citenamefont {{Choi}}, \citenamefont {{Yao}},\
  and\ \citenamefont {{Lukin}}}]{ChoiMetrology}%
  \BibitemOpen
  \bibfield  {author} {\bibinfo {author} {\bibfnamefont {S.}~\bibnamefont
  {{Choi}}}, \bibinfo {author} {\bibfnamefont {N.~Y.}\ \bibnamefont {{Yao}}}, \
  and\ \bibinfo {author} {\bibfnamefont {M.~D.}\ \bibnamefont {{Lukin}}},\
  }\bibfield  {title} {\enquote {\bibinfo {title} {{Quantum metrology based on
  strongly correlated matter}},}\ }\href@noop {} {\bibfield  {journal}
  {\bibinfo  {journal} {ArXiv e-prints}\ } (\bibinfo {year}
  {2018}{\natexlab{a}})},\ \Eprint {http://arxiv.org/abs/1801.00042}
  {arXiv:1801.00042 [quant-ph]} \BibitemShut {NoStop}%
\bibitem [{\citenamefont {Aidelsburger}\ \emph {et~al.}(2013)\citenamefont
  {Aidelsburger}, \citenamefont {Atala}, \citenamefont {Lohse}, \citenamefont
  {Barreiro}, \citenamefont {Paredes},\ and\ \citenamefont
  {Bloch}}]{PhysRevLett.111.185301}%
  \BibitemOpen
  \bibfield  {author} {\bibinfo {author} {\bibfnamefont {M.}~\bibnamefont
  {Aidelsburger}}, \bibinfo {author} {\bibfnamefont {M.}~\bibnamefont {Atala}},
  \bibinfo {author} {\bibfnamefont {M.}~\bibnamefont {Lohse}}, \bibinfo
  {author} {\bibfnamefont {J.~T.}\ \bibnamefont {Barreiro}}, \bibinfo {author}
  {\bibfnamefont {B.}~\bibnamefont {Paredes}}, \ and\ \bibinfo {author}
  {\bibfnamefont {I.}~\bibnamefont {Bloch}},\ }\bibfield  {title} {\enquote
  {\bibinfo {title} {Realization of the hofstadter hamiltonian with ultracold
  atoms in optical lattices},}\ }\href {\doibase
  10.1103/PhysRevLett.111.185301} {\bibfield  {journal} {\bibinfo  {journal}
  {Phys. Rev. Lett.}\ }\textbf {\bibinfo {volume} {111}},\ \bibinfo {pages}
  {185301} (\bibinfo {year} {2013})}\BibitemShut {NoStop}%
\bibitem [{\citenamefont {Aidelsburger}\ \emph {et~al.}(2014)\citenamefont
  {Aidelsburger}, \citenamefont {Lohse}, \citenamefont {Schweizer},
  \citenamefont {Atala}, \citenamefont {Barreiro}, \citenamefont
  {Nascimb{\`e}ne}, \citenamefont {Cooper}, \citenamefont {Bloch},\ and\
  \citenamefont {Goldman}}]{Aidelsburger2014}%
  \BibitemOpen
  \bibfield  {author} {\bibinfo {author} {\bibfnamefont {M.}~\bibnamefont
  {Aidelsburger}}, \bibinfo {author} {\bibfnamefont {M.}~\bibnamefont {Lohse}},
  \bibinfo {author} {\bibfnamefont {C.}~\bibnamefont {Schweizer}}, \bibinfo
  {author} {\bibfnamefont {M.}~\bibnamefont {Atala}}, \bibinfo {author}
  {\bibfnamefont {J.T.}\ \bibnamefont {Barreiro}}, \bibinfo {author}
  {\bibfnamefont {S.}~\bibnamefont {Nascimb{\`e}ne}}, \bibinfo {author}
  {\bibfnamefont {N.~R.}\ \bibnamefont {Cooper}}, \bibinfo {author}
  {\bibfnamefont {I.}~\bibnamefont {Bloch}}, \ and\ \bibinfo {author}
  {\bibfnamefont {N.}~\bibnamefont {Goldman}},\ }\bibfield  {title} {\enquote
  {\bibinfo {title} {Measuring the chern number of hofstadter bands with
  ultracold bosonic atoms},}\ }\href {http://dx.doi.org/10.1038/nphys3171}
  {\bibfield  {journal} {\bibinfo  {journal} {Nature Physics}\ }\textbf
  {\bibinfo {volume} {11}},\ \bibinfo {pages} {162 EP --} (\bibinfo {year}
  {2014})}\BibitemShut {NoStop}%
\bibitem [{\citenamefont {Islam}\ \emph {et~al.}(2015)\citenamefont {Islam},
  \citenamefont {Ma}, \citenamefont {Preiss}, \citenamefont {Eric~Tai},
  \citenamefont {Lukin}, \citenamefont {Rispoli},\ and\ \citenamefont
  {Greiner}}]{Islam2015}%
  \BibitemOpen
  \bibfield  {author} {\bibinfo {author} {\bibfnamefont {Rajibul}\ \bibnamefont
  {Islam}}, \bibinfo {author} {\bibfnamefont {Ruichao}\ \bibnamefont {Ma}},
  \bibinfo {author} {\bibfnamefont {Philipp~M.}\ \bibnamefont {Preiss}},
  \bibinfo {author} {\bibfnamefont {M.}~\bibnamefont {Eric~Tai}}, \bibinfo
  {author} {\bibfnamefont {Alexander}\ \bibnamefont {Lukin}}, \bibinfo {author}
  {\bibfnamefont {Matthew}\ \bibnamefont {Rispoli}}, \ and\ \bibinfo {author}
  {\bibfnamefont {Markus}\ \bibnamefont {Greiner}},\ }\bibfield  {title}
  {\enquote {\bibinfo {title} {Measuring entanglement entropy in a quantum
  many-body system},}\ }\href {http://dx.doi.org/10.1038/nature15750}
  {\bibfield  {journal} {\bibinfo  {journal} {Nature}\ }\textbf {\bibinfo
  {volume} {528}},\ \bibinfo {pages} {77 EP --} (\bibinfo {year} {2015})},\
  \bibinfo {note} {article}\BibitemShut {NoStop}%
\bibitem [{\citenamefont {Choi}\ \emph {et~al.}(2016)\citenamefont {Choi},
  \citenamefont {Hild}, \citenamefont {Zeiher}, \citenamefont {Schau{\ss}},
  \citenamefont {Rubio-Abadal}, \citenamefont {Yefsah}, \citenamefont
  {Khemani}, \citenamefont {Huse}, \citenamefont {Bloch},\ and\ \citenamefont
  {Gross}}]{Bloch16}%
  \BibitemOpen
  \bibfield  {author} {\bibinfo {author} {\bibfnamefont {Jae-yoon}\
  \bibnamefont {Choi}}, \bibinfo {author} {\bibfnamefont {Sebastian}\
  \bibnamefont {Hild}}, \bibinfo {author} {\bibfnamefont {Johannes}\
  \bibnamefont {Zeiher}}, \bibinfo {author} {\bibfnamefont {Peter}\
  \bibnamefont {Schau{\ss}}}, \bibinfo {author} {\bibfnamefont {Antonio}\
  \bibnamefont {Rubio-Abadal}}, \bibinfo {author} {\bibfnamefont {Tarik}\
  \bibnamefont {Yefsah}}, \bibinfo {author} {\bibfnamefont {Vedika}\
  \bibnamefont {Khemani}}, \bibinfo {author} {\bibfnamefont {David~A.}\
  \bibnamefont {Huse}}, \bibinfo {author} {\bibfnamefont {Immanuel}\
  \bibnamefont {Bloch}}, \ and\ \bibinfo {author} {\bibfnamefont {Christian}\
  \bibnamefont {Gross}},\ }\bibfield  {title} {\enquote {\bibinfo {title}
  {Exploring the many-body localization transition in two dimensions},}\ }\href
  {\doibase 10.1126/science.aaf8834} {\bibfield  {journal} {\bibinfo  {journal}
  {Science}\ }\textbf {\bibinfo {volume} {352}},\ \bibinfo {pages} {1547--1552}
  (\bibinfo {year} {2016})}\BibitemShut {NoStop}%
\bibitem [{\citenamefont {Smith}\ \emph {et~al.}(2016)\citenamefont {Smith},
  \citenamefont {Lee}, \citenamefont {Richerme}, \citenamefont {Neyenhuis},
  \citenamefont {Hess}, \citenamefont {Hauke}, \citenamefont {Heyl},
  \citenamefont {Huse},\ and\ \citenamefont {Monroe}}]{Smith2016}%
  \BibitemOpen
  \bibfield  {author} {\bibinfo {author} {\bibfnamefont {J.}~\bibnamefont
  {Smith}}, \bibinfo {author} {\bibfnamefont {A.}~\bibnamefont {Lee}}, \bibinfo
  {author} {\bibfnamefont {P.}~\bibnamefont {Richerme}}, \bibinfo {author}
  {\bibfnamefont {B.}~\bibnamefont {Neyenhuis}}, \bibinfo {author}
  {\bibfnamefont {P.~W.}\ \bibnamefont {Hess}}, \bibinfo {author}
  {\bibfnamefont {P.}~\bibnamefont {Hauke}}, \bibinfo {author} {\bibfnamefont
  {M.}~\bibnamefont {Heyl}}, \bibinfo {author} {\bibfnamefont {D.~A.}\
  \bibnamefont {Huse}}, \ and\ \bibinfo {author} {\bibfnamefont
  {C.}~\bibnamefont {Monroe}},\ }\bibfield  {title} {\enquote {\bibinfo {title}
  {Many-body localization in a quantum simulator with programmable random
  disorder},}\ }\href {http://dx.doi.org/10.1038/nphys3783} {\bibfield
  {journal} {\bibinfo  {journal} {Nature Physics}\ }\textbf {\bibinfo {volume}
  {12}},\ \bibinfo {pages} {907 EP --} (\bibinfo {year} {2016})}\BibitemShut
  {NoStop}%
\bibitem [{\citenamefont {Choi}\ \emph {et~al.}(2017)\citenamefont {Choi},
  \citenamefont {Choi}, \citenamefont {Landig}, \citenamefont {Kucsko},
  \citenamefont {Zhou}, \citenamefont {Isoya}, \citenamefont {Jelezko},
  \citenamefont {Onoda}, \citenamefont {Sumiya}, \citenamefont {Khemani},
  \citenamefont {von Keyserlingk}, \citenamefont {Yao}, \citenamefont
  {Demler},\ and\ \citenamefont {Lukin}}]{Choi16DTC}%
  \BibitemOpen
  \bibfield  {author} {\bibinfo {author} {\bibfnamefont {Soonwon}\ \bibnamefont
  {Choi}}, \bibinfo {author} {\bibfnamefont {Joonhee}\ \bibnamefont {Choi}},
  \bibinfo {author} {\bibfnamefont {Renate}\ \bibnamefont {Landig}}, \bibinfo
  {author} {\bibfnamefont {Georg}\ \bibnamefont {Kucsko}}, \bibinfo {author}
  {\bibfnamefont {Hengyun}\ \bibnamefont {Zhou}}, \bibinfo {author}
  {\bibfnamefont {Junichi}\ \bibnamefont {Isoya}}, \bibinfo {author}
  {\bibfnamefont {Fedor}\ \bibnamefont {Jelezko}}, \bibinfo {author}
  {\bibfnamefont {Shinobu}\ \bibnamefont {Onoda}}, \bibinfo {author}
  {\bibfnamefont {Hitoshi}\ \bibnamefont {Sumiya}}, \bibinfo {author}
  {\bibfnamefont {Vedika}\ \bibnamefont {Khemani}}, \bibinfo {author}
  {\bibfnamefont {Curt}\ \bibnamefont {von Keyserlingk}}, \bibinfo {author}
  {\bibfnamefont {Norman~Y.}\ \bibnamefont {Yao}}, \bibinfo {author}
  {\bibfnamefont {Eugene}\ \bibnamefont {Demler}}, \ and\ \bibinfo {author}
  {\bibfnamefont {Mikhail~D.}\ \bibnamefont {Lukin}},\ }\bibfield  {title}
  {\enquote {\bibinfo {title} {Observation of discrete time-crystalline order
  in a disordered dipolar many-body system},}\ }\href
  {http://dx.doi.org/10.1038/nature21426} {\bibfield  {journal} {\bibinfo
  {journal} {Nature}\ }\textbf {\bibinfo {volume} {543}},\ \bibinfo {pages}
  {221--225} (\bibinfo {year} {2017})}\BibitemShut {NoStop}%
\bibitem [{\citenamefont {Zhang}\ \emph
  {et~al.}(2017{\natexlab{b}})\citenamefont {Zhang}, \citenamefont {Hess},
  \citenamefont {Kyprianidis}, \citenamefont {Becker}, \citenamefont {Lee},
  \citenamefont {Smith}, \citenamefont {Pagano}, \citenamefont {Potirniche},
  \citenamefont {Potter}, \citenamefont {Vishwanath}, \citenamefont {Yao},\
  and\ \citenamefont {Monroe}}]{Zhang2017}%
  \BibitemOpen
  \bibfield  {author} {\bibinfo {author} {\bibfnamefont {J.}~\bibnamefont
  {Zhang}}, \bibinfo {author} {\bibfnamefont {P.~W.}\ \bibnamefont {Hess}},
  \bibinfo {author} {\bibfnamefont {A.}~\bibnamefont {Kyprianidis}}, \bibinfo
  {author} {\bibfnamefont {P.}~\bibnamefont {Becker}}, \bibinfo {author}
  {\bibfnamefont {A.}~\bibnamefont {Lee}}, \bibinfo {author} {\bibfnamefont
  {J.}~\bibnamefont {Smith}}, \bibinfo {author} {\bibfnamefont
  {G.}~\bibnamefont {Pagano}}, \bibinfo {author} {\bibfnamefont {I.-D.}\
  \bibnamefont {Potirniche}}, \bibinfo {author} {\bibfnamefont {A.~C.}\
  \bibnamefont {Potter}}, \bibinfo {author} {\bibfnamefont {A.}~\bibnamefont
  {Vishwanath}}, \bibinfo {author} {\bibfnamefont {N.~Y.}\ \bibnamefont {Yao}},
  \ and\ \bibinfo {author} {\bibfnamefont {C.}~\bibnamefont {Monroe}},\
  }\bibfield  {title} {\enquote {\bibinfo {title} {Observation of a discrete
  time crystal},}\ }\href {http://dx.doi.org/10.1038/nature21413} {\bibfield
  {journal} {\bibinfo  {journal} {Nature}\ }\textbf {\bibinfo {volume} {543}},\
  \bibinfo {pages} {217--220} (\bibinfo {year}
  {2017}{\natexlab{b}})}\BibitemShut {NoStop}%
\bibitem [{\citenamefont {{Choi}}\ \emph
  {et~al.}(2018{\natexlab{b}})\citenamefont {{Choi}}, \citenamefont {{Zhou}},
  \citenamefont {{Choi}}, \citenamefont {{Landig}}, \citenamefont {{Ho}},
  \citenamefont {{Isoya}}, \citenamefont {{Jelezko}}, \citenamefont {{Onoda}},
  \citenamefont {{Sumiya}}, \citenamefont {{Abanin}},\ and\ \citenamefont
  {{Lukin}}}]{2018arXiv180610169C}%
  \BibitemOpen
  \bibfield  {author} {\bibinfo {author} {\bibfnamefont {J.}~\bibnamefont
  {{Choi}}}, \bibinfo {author} {\bibfnamefont {H.}~\bibnamefont {{Zhou}}},
  \bibinfo {author} {\bibfnamefont {S.}~\bibnamefont {{Choi}}}, \bibinfo
  {author} {\bibfnamefont {R.}~\bibnamefont {{Landig}}}, \bibinfo {author}
  {\bibfnamefont {W.~W.}\ \bibnamefont {{Ho}}}, \bibinfo {author}
  {\bibfnamefont {J.}~\bibnamefont {{Isoya}}}, \bibinfo {author} {\bibfnamefont
  {F.}~\bibnamefont {{Jelezko}}}, \bibinfo {author} {\bibfnamefont
  {S.}~\bibnamefont {{Onoda}}}, \bibinfo {author} {\bibfnamefont
  {H.}~\bibnamefont {{Sumiya}}}, \bibinfo {author} {\bibfnamefont {D.~A.}\
  \bibnamefont {{Abanin}}}, \ and\ \bibinfo {author} {\bibfnamefont {M.~D.}\
  \bibnamefont {{Lukin}}},\ }\bibfield  {title} {\enquote {\bibinfo {title}
  {{Probing quantum thermalization of a disordered dipolar spin ensemble with
  discrete time-crystalline order}},}\ }\href@noop {} {\bibfield  {journal}
  {\bibinfo  {journal} {ArXiv e-prints}\ } (\bibinfo {year}
  {2018}{\natexlab{b}})},\ \Eprint {http://arxiv.org/abs/1806.10169}
  {arXiv:1806.10169 [quant-ph]} \BibitemShut {NoStop}%
\bibitem [{\citenamefont {Ho}\ and\ \citenamefont
  {Hsieh}(2019)}]{2018arXiv180300026H}%
  \BibitemOpen
  \bibfield  {author} {\bibinfo {author} {\bibfnamefont {Wen~Wei}\ \bibnamefont
  {Ho}}\ and\ \bibinfo {author} {\bibfnamefont {Timothy~H.}\ \bibnamefont
  {Hsieh}},\ }\bibfield  {title} {\enquote {\bibinfo {title} {{Efficient
  variational simulation of non-trivial quantum states}},}\ }\href {\doibase
  10.21468/SciPostPhys.6.3.029} {\bibfield  {journal} {\bibinfo  {journal}
  {SciPost Phys.}\ }\textbf {\bibinfo {volume} {6}},\ \bibinfo {pages} {29}
  (\bibinfo {year} {2019})}\BibitemShut {NoStop}%
\bibitem [{\citenamefont {{Farhi}}\ \emph {et~al.}(2014)\citenamefont
  {{Farhi}}, \citenamefont {{Goldstone}},\ and\ \citenamefont
  {{Gutmann}}}]{QAOA1}%
  \BibitemOpen
  \bibfield  {author} {\bibinfo {author} {\bibfnamefont {E.}~\bibnamefont
  {{Farhi}}}, \bibinfo {author} {\bibfnamefont {J.}~\bibnamefont
  {{Goldstone}}}, \ and\ \bibinfo {author} {\bibfnamefont {S.}~\bibnamefont
  {{Gutmann}}},\ }\bibfield  {title} {\enquote {\bibinfo {title} {{A Quantum
  Approximate Optimization Algorithm}},}\ }\href@noop {} {\bibfield  {journal}
  {\bibinfo  {journal} {ArXiv e-prints}\ } (\bibinfo {year} {2014})},\ \Eprint
  {http://arxiv.org/abs/1411.4028} {arXiv:1411.4028 [quant-ph]} \BibitemShut
  {NoStop}%
\bibitem [{\citenamefont {{Farhi}}\ and\ \citenamefont
  {{Harrow}}(2016)}]{QAOA2}%
  \BibitemOpen
  \bibfield  {author} {\bibinfo {author} {\bibfnamefont {E.}~\bibnamefont
  {{Farhi}}}\ and\ \bibinfo {author} {\bibfnamefont {A.~W}\ \bibnamefont
  {{Harrow}}},\ }\bibfield  {title} {\enquote {\bibinfo {title} {{Quantum
  Supremacy through the Quantum Approximate Optimization Algorithm}},}\
  }\href@noop {} {\bibfield  {journal} {\bibinfo  {journal} {ArXiv e-prints}\ }
  (\bibinfo {year} {2016})},\ \Eprint {http://arxiv.org/abs/1602.07674}
  {arXiv:1602.07674 [quant-ph]} \BibitemShut {NoStop}%
\bibitem [{\citenamefont {Wecker}\ \emph {et~al.}(2015)\citenamefont {Wecker},
  \citenamefont {Hastings},\ and\ \citenamefont {Troyer}}]{variational}%
  \BibitemOpen
  \bibfield  {author} {\bibinfo {author} {\bibfnamefont {Dave}\ \bibnamefont
  {Wecker}}, \bibinfo {author} {\bibfnamefont {Matthew~B.}\ \bibnamefont
  {Hastings}}, \ and\ \bibinfo {author} {\bibfnamefont {Matthias}\ \bibnamefont
  {Troyer}},\ }\bibfield  {title} {\enquote {\bibinfo {title} {Progress towards
  practical quantum variational algorithms},}\ }\href {\doibase
  10.1103/PhysRevA.92.042303} {\bibfield  {journal} {\bibinfo  {journal} {Phys.
  Rev. A}\ }\textbf {\bibinfo {volume} {92}},\ \bibinfo {pages} {042303}
  (\bibinfo {year} {2015})}\BibitemShut {NoStop}%
\bibitem [{\citenamefont {Peruzzo}\ \emph {et~al.}(2014)\citenamefont
  {Peruzzo}, \citenamefont {McClean}, \citenamefont {Shadbolt}, \citenamefont
  {Yung}, \citenamefont {Zhou}, \citenamefont {Love}, \citenamefont
  {Aspuru-Guzik},\ and\ \citenamefont {O'Brien}}]{VQE}%
  \BibitemOpen
  \bibfield  {author} {\bibinfo {author} {\bibfnamefont {Alberto}\ \bibnamefont
  {Peruzzo}}, \bibinfo {author} {\bibfnamefont {Jarrod}\ \bibnamefont
  {McClean}}, \bibinfo {author} {\bibfnamefont {Peter}\ \bibnamefont
  {Shadbolt}}, \bibinfo {author} {\bibfnamefont {Man-Hong}\ \bibnamefont
  {Yung}}, \bibinfo {author} {\bibfnamefont {Xiao-Qi}\ \bibnamefont {Zhou}},
  \bibinfo {author} {\bibfnamefont {Peter~J.}\ \bibnamefont {Love}}, \bibinfo
  {author} {\bibfnamefont {Al{\'a}n}\ \bibnamefont {Aspuru-Guzik}}, \ and\
  \bibinfo {author} {\bibfnamefont {Jeremy~L.}\ \bibnamefont {O'Brien}},\
  }\bibfield  {title} {\enquote {\bibinfo {title} {A variational eigenvalue
  solver on a photonic quantum processor},}\ }\href
  {https://doi.org/10.1038/ncomms5213} {\bibfield  {journal} {\bibinfo
  {journal} {Nature Communications}\ }\textbf {\bibinfo {volume} {5}},\
  \bibinfo {pages} {4213 EP --} (\bibinfo {year} {2014})}\BibitemShut {NoStop}%
\bibitem [{\citenamefont {Demirplak}\ and\ \citenamefont
  {Rice}(2003)}]{doi:10.1021/jp030708a}%
  \BibitemOpen
  \bibfield  {author} {\bibinfo {author} {\bibfnamefont {Mustafa}\ \bibnamefont
  {Demirplak}}\ and\ \bibinfo {author} {\bibfnamefont {Stuart~A.}\ \bibnamefont
  {Rice}},\ }\bibfield  {title} {\enquote {\bibinfo {title} {Adiabatic
  population transfer with control fields},}\ }\href {\doibase
  10.1021/jp030708a} {\bibfield  {journal} {\bibinfo  {journal} {The Journal of
  Physical Chemistry A}\ }\textbf {\bibinfo {volume} {107}},\ \bibinfo {pages}
  {9937--9945} (\bibinfo {year} {2003})},\ \Eprint
  {http://arxiv.org/abs/https://doi.org/10.1021/jp030708a}
  {https://doi.org/10.1021/jp030708a} \BibitemShut {NoStop}%
\bibitem [{\citenamefont {Demirplak}\ and\ \citenamefont
  {Rice}(2005)}]{doi:10.1021/jp040647w}%
  \BibitemOpen
  \bibfield  {author} {\bibinfo {author} {\bibfnamefont {Mustafa}\ \bibnamefont
  {Demirplak}}\ and\ \bibinfo {author} {\bibfnamefont {Stuart~A.}\ \bibnamefont
  {Rice}},\ }\bibfield  {title} {\enquote {\bibinfo {title} {Assisted adiabatic
  passage revisited},}\ }\href {\doibase 10.1021/jp040647w} {\bibfield
  {journal} {\bibinfo  {journal} {The Journal of Physical Chemistry B}\
  }\textbf {\bibinfo {volume} {109}},\ \bibinfo {pages} {6838--6844} (\bibinfo
  {year} {2005})},\ \bibinfo {note} {pMID: 16851769},\ \Eprint
  {http://arxiv.org/abs/https://doi.org/10.1021/jp040647w}
  {https://doi.org/10.1021/jp040647w} \BibitemShut {NoStop}%
\bibitem [{\citenamefont {Berry}(2009)}]{1751-8121-42-36-365303}%
  \BibitemOpen
  \bibfield  {author} {\bibinfo {author} {\bibfnamefont {M~V}\ \bibnamefont
  {Berry}},\ }\bibfield  {title} {\enquote {\bibinfo {title} {Transitionless
  quantum driving},}\ }\href {http://stacks.iop.org/1751-8121/42/i=36/a=365303}
  {\bibfield  {journal} {\bibinfo  {journal} {Journal of Physics A:
  Mathematical and Theoretical}\ }\textbf {\bibinfo {volume} {42}},\ \bibinfo
  {pages} {365303} (\bibinfo {year} {2009})}\BibitemShut {NoStop}%
\bibitem [{\citenamefont {{Agarwal}}\ \emph {et~al.}(2017)\citenamefont
  {{Agarwal}}, \citenamefont {{Bhatt}},\ and\ \citenamefont
  {{Sondhi}}}]{QuenchedPrep}%
  \BibitemOpen
  \bibfield  {author} {\bibinfo {author} {\bibfnamefont {K.}~\bibnamefont
  {{Agarwal}}}, \bibinfo {author} {\bibfnamefont {R.~N.}\ \bibnamefont
  {{Bhatt}}}, \ and\ \bibinfo {author} {\bibfnamefont {S.~L.}\ \bibnamefont
  {{Sondhi}}},\ }\bibfield  {title} {\enquote {\bibinfo {title} {{Fast
  preparation of critical ground states using superluminal fronts}},}\
  }\href@noop {} {\bibfield  {journal} {\bibinfo  {journal} {ArXiv e-prints}\ }
  (\bibinfo {year} {2017})},\ \Eprint {http://arxiv.org/abs/1710.09840}
  {arXiv:1710.09840 [cond-mat.quant-gas]} \BibitemShut {NoStop}%
\bibitem [{\citenamefont {Sels}\ and\ \citenamefont
  {Polkovnikov}(2017)}]{Sels201619826}%
  \BibitemOpen
  \bibfield  {author} {\bibinfo {author} {\bibfnamefont {Dries}\ \bibnamefont
  {Sels}}\ and\ \bibinfo {author} {\bibfnamefont {Anatoli}\ \bibnamefont
  {Polkovnikov}},\ }\bibfield  {title} {\enquote {\bibinfo {title} {Minimizing
  irreversible losses in quantum systems by local counterdiabatic driving},}\
  }\href {\doibase 10.1073/pnas.1619826114} {\bibfield  {journal} {\bibinfo
  {journal} {Proceedings of the National Academy of Sciences}\ } (\bibinfo
  {year} {2017}),\ 10.1073/pnas.1619826114}\BibitemShut {NoStop}%
\bibitem [{\citenamefont {{Farhi}}\ \emph {et~al.}(2000)\citenamefont
  {{Farhi}}, \citenamefont {{Goldstone}}, \citenamefont {{Gutmann}},\ and\
  \citenamefont {{Sipser}}}]{QAA1}%
  \BibitemOpen
  \bibfield  {author} {\bibinfo {author} {\bibfnamefont {E.}~\bibnamefont
  {{Farhi}}}, \bibinfo {author} {\bibfnamefont {J.}~\bibnamefont
  {{Goldstone}}}, \bibinfo {author} {\bibfnamefont {S.}~\bibnamefont
  {{Gutmann}}}, \ and\ \bibinfo {author} {\bibfnamefont {M.}~\bibnamefont
  {{Sipser}}},\ }\bibfield  {title} {\enquote {\bibinfo {title} {{Quantum
  Computation by Adiabatic Evolution}},}\ }\href@noop {} {\bibfield  {journal}
  {\bibinfo  {journal} {eprint arXiv:quant-ph/0001106}\ } (\bibinfo {year}
  {2000})},\ \Eprint {http://arxiv.org/abs/quant-ph/0001106} {quant-ph/0001106}
  \BibitemShut {NoStop}%
\bibitem [{\citenamefont {Farhi}\ \emph {et~al.}(2001)\citenamefont {Farhi},
  \citenamefont {Goldstone}, \citenamefont {Gutmann}, \citenamefont {Lapan},
  \citenamefont {Lundgren},\ and\ \citenamefont {Preda}}]{QAA2}%
  \BibitemOpen
  \bibfield  {author} {\bibinfo {author} {\bibfnamefont {Edward}\ \bibnamefont
  {Farhi}}, \bibinfo {author} {\bibfnamefont {Jeffrey}\ \bibnamefont
  {Goldstone}}, \bibinfo {author} {\bibfnamefont {Sam}\ \bibnamefont
  {Gutmann}}, \bibinfo {author} {\bibfnamefont {Joshua}\ \bibnamefont {Lapan}},
  \bibinfo {author} {\bibfnamefont {Andrew}\ \bibnamefont {Lundgren}}, \ and\
  \bibinfo {author} {\bibfnamefont {Daniel}\ \bibnamefont {Preda}},\ }\bibfield
   {title} {\enquote {\bibinfo {title} {A quantum adiabatic evolution algorithm
  applied to random instances of an np-complete problem},}\ }\href {\doibase
  10.1126/science.1057726} {\bibfield  {journal} {\bibinfo  {journal}
  {Science}\ }\textbf {\bibinfo {volume} {292}},\ \bibinfo {pages} {472--475}
  (\bibinfo {year} {2001})}\BibitemShut {NoStop}%
\bibitem [{\citenamefont {Pontryagin}(1987)}]{optimalControl1}%
  \BibitemOpen
  \bibfield  {author} {\bibinfo {author} {\bibfnamefont {L.S.}\ \bibnamefont
  {Pontryagin}},\ }\href@noop {} {\emph {\bibinfo {title} {Mathematical Theory
  of Optimal Processes}}}\ (\bibinfo {year} {1987})\BibitemShut {NoStop}%
\bibitem [{\citenamefont {Stengel}(1994)}]{optimalControl2}%
  \BibitemOpen
  \bibfield  {author} {\bibinfo {author} {\bibfnamefont {Robert~F.}\
  \bibnamefont {Stengel}},\ }\href@noop {} {\emph {\bibinfo {title} {Optimal
  Control and Estimation}}}\ (\bibinfo {year} {1994})\BibitemShut {NoStop}%
\bibitem [{\citenamefont {Brif}\ \emph {et~al.}(2014)\citenamefont {Brif},
  \citenamefont {Grace}, \citenamefont {Sarovar},\ and\ \citenamefont
  {Young}}]{OptimalControl3}%
  \BibitemOpen
  \bibfield  {author} {\bibinfo {author} {\bibfnamefont {Constantin}\
  \bibnamefont {Brif}}, \bibinfo {author} {\bibfnamefont {Matthew~D}\
  \bibnamefont {Grace}}, \bibinfo {author} {\bibfnamefont {Mohan}\ \bibnamefont
  {Sarovar}}, \ and\ \bibinfo {author} {\bibfnamefont {Kevin~C}\ \bibnamefont
  {Young}},\ }\bibfield  {title} {\enquote {\bibinfo {title} {Exploring
  adiabatic quantum trajectories via optimal control},}\ }\href
  {http://stacks.iop.org/1367-2630/16/i=6/a=065013} {\bibfield  {journal}
  {\bibinfo  {journal} {New Journal of Physics}\ }\textbf {\bibinfo {volume}
  {16}},\ \bibinfo {pages} {065013} (\bibinfo {year} {2014})}\BibitemShut
  {NoStop}%
\bibitem [{\citenamefont {Yang}\ \emph {et~al.}(2017)\citenamefont {Yang},
  \citenamefont {Rahmani}, \citenamefont {Shabani}, \citenamefont {Neven},\
  and\ \citenamefont {Chamon}}]{PhysRevX.7.021027}%
  \BibitemOpen
  \bibfield  {author} {\bibinfo {author} {\bibfnamefont {Zhi-Cheng}\
  \bibnamefont {Yang}}, \bibinfo {author} {\bibfnamefont {Armin}\ \bibnamefont
  {Rahmani}}, \bibinfo {author} {\bibfnamefont {Alireza}\ \bibnamefont
  {Shabani}}, \bibinfo {author} {\bibfnamefont {Hartmut}\ \bibnamefont
  {Neven}}, \ and\ \bibinfo {author} {\bibfnamefont {Claudio}\ \bibnamefont
  {Chamon}},\ }\bibfield  {title} {\enquote {\bibinfo {title} {Optimizing
  variational quantum algorithms using pontryagin's minimum principle},}\
  }\href {\doibase 10.1103/PhysRevX.7.021027} {\bibfield  {journal} {\bibinfo
  {journal} {Phys. Rev. X}\ }\textbf {\bibinfo {volume} {7}},\ \bibinfo {pages}
  {021027} (\bibinfo {year} {2017})}\BibitemShut {NoStop}%
\bibitem [{\citenamefont {Lieb}\ and\ \citenamefont
  {Robinson}(1972)}]{lieb1972}%
  \BibitemOpen
  \bibfield  {author} {\bibinfo {author} {\bibfnamefont {Elliott~H.}\
  \bibnamefont {Lieb}}\ and\ \bibinfo {author} {\bibfnamefont {Derek~W.}\
  \bibnamefont {Robinson}},\ }\bibfield  {title} {\enquote {\bibinfo {title}
  {The finite group velocity of quantum spin systems},}\ }\href
  {https://projecteuclid.org:443/euclid.cmp/1103858407} {\bibfield  {journal}
  {\bibinfo  {journal} {Comm. Math. Phys.}\ }\textbf {\bibinfo {volume} {28}},\
  \bibinfo {pages} {251--257} (\bibinfo {year} {1972})}\BibitemShut {NoStop}%
\bibitem [{\citenamefont {Nielsen}(2006)}]{nielsen}%
  \BibitemOpen
  \bibfield  {author} {\bibinfo {author} {\bibfnamefont {Michael~A.}\
  \bibnamefont {Nielsen}},\ }\bibfield  {title} {\enquote {\bibinfo {title} {A
  geometric approach to quantum circuit lower bounds},}\ }\href
  {http://dl.acm.org/citation.cfm?id=2011686.2011688} {\bibfield  {journal}
  {\bibinfo  {journal} {Quantum Info. Comput.}\ }\textbf {\bibinfo {volume}
  {6}},\ \bibinfo {pages} {213--262} (\bibinfo {year} {2006})}\BibitemShut
  {NoStop}%
\bibitem [{\citenamefont {Bravyi}\ \emph {et~al.}(2006)\citenamefont {Bravyi},
  \citenamefont {Hastings},\ and\ \citenamefont
  {Verstraete}}]{PhysRevLett.97.050401}%
  \BibitemOpen
  \bibfield  {author} {\bibinfo {author} {\bibfnamefont {S.}~\bibnamefont
  {Bravyi}}, \bibinfo {author} {\bibfnamefont {M.~B.}\ \bibnamefont
  {Hastings}}, \ and\ \bibinfo {author} {\bibfnamefont {F.}~\bibnamefont
  {Verstraete}},\ }\bibfield  {title} {\enquote {\bibinfo {title}
  {Lieb-robinson bounds and the generation of correlations and topological
  quantum order},}\ }\href {\doibase 10.1103/PhysRevLett.97.050401} {\bibfield
  {journal} {\bibinfo  {journal} {Phys. Rev. Lett.}\ }\textbf {\bibinfo
  {volume} {97}},\ \bibinfo {pages} {050401} (\bibinfo {year}
  {2006})}\BibitemShut {NoStop}%
\bibitem [{\citenamefont {{Hastings}}(2010)}]{HastingsLR}%
  \BibitemOpen
  \bibfield  {author} {\bibinfo {author} {\bibfnamefont {M.~B.}\ \bibnamefont
  {{Hastings}}},\ }\bibfield  {title} {\enquote {\bibinfo {title} {{Locality in
  Quantum Systems}},}\ }\href@noop {} {\bibfield  {journal} {\bibinfo
  {journal} {ArXiv e-prints}\ } (\bibinfo {year} {2010})},\ \Eprint
  {http://arxiv.org/abs/1008.5137} {arXiv:1008.5137 [math-ph]} \BibitemShut
  {NoStop}%
\bibitem [{\citenamefont {Richerme}\ \emph {et~al.}(2014)\citenamefont
  {Richerme}, \citenamefont {Gong}, \citenamefont {Lee}, \citenamefont {Senko},
  \citenamefont {Smith}, \citenamefont {Foss-Feig}, \citenamefont {Michalakis},
  \citenamefont {Gorshkov},\ and\ \citenamefont {Monroe}}]{Richerme2014}%
  \BibitemOpen
  \bibfield  {author} {\bibinfo {author} {\bibfnamefont {Philip}\ \bibnamefont
  {Richerme}}, \bibinfo {author} {\bibfnamefont {Zhe-Xuan}\ \bibnamefont
  {Gong}}, \bibinfo {author} {\bibfnamefont {Aaron}\ \bibnamefont {Lee}},
  \bibinfo {author} {\bibfnamefont {Crystal}\ \bibnamefont {Senko}}, \bibinfo
  {author} {\bibfnamefont {Jacob}\ \bibnamefont {Smith}}, \bibinfo {author}
  {\bibfnamefont {Michael}\ \bibnamefont {Foss-Feig}}, \bibinfo {author}
  {\bibfnamefont {Spyridon}\ \bibnamefont {Michalakis}}, \bibinfo {author}
  {\bibfnamefont {Alexey~V.}\ \bibnamefont {Gorshkov}}, \ and\ \bibinfo
  {author} {\bibfnamefont {Christopher}\ \bibnamefont {Monroe}},\ }\bibfield
  {title} {\enquote {\bibinfo {title} {Non-local propagation of correlations in
  quantum systems with long-range interactions},}\ }\href
  {http://dx.doi.org/10.1038/nature13450} {\bibfield  {journal} {\bibinfo
  {journal} {Nature}\ }\textbf {\bibinfo {volume} {511}},\ \bibinfo {pages}
  {198 EP --} (\bibinfo {year} {2014})}\BibitemShut {NoStop}%
\bibitem [{\citenamefont {Matsuta}\ \emph {et~al.}(2017)\citenamefont
  {Matsuta}, \citenamefont {Koma},\ and\ \citenamefont {Nakamura}}]{LR_LR1}%
  \BibitemOpen
  \bibfield  {author} {\bibinfo {author} {\bibfnamefont {Takuro}\ \bibnamefont
  {Matsuta}}, \bibinfo {author} {\bibfnamefont {Tohru}\ \bibnamefont {Koma}}, \
  and\ \bibinfo {author} {\bibfnamefont {Shu}\ \bibnamefont {Nakamura}},\
  }\bibfield  {title} {\enquote {\bibinfo {title} {Improving the lieb--robinson
  bound for long-range interactions},}\ }\href {\doibase
  10.1007/s00023-016-0526-1} {\bibfield  {journal} {\bibinfo  {journal}
  {Annales Henri Poincar{\'e}}\ }\textbf {\bibinfo {volume} {18}},\ \bibinfo
  {pages} {519--528} (\bibinfo {year} {2017})}\BibitemShut {NoStop}%
\bibitem [{\citenamefont {Foss-Feig}\ \emph {et~al.}(2015)\citenamefont
  {Foss-Feig}, \citenamefont {Gong}, \citenamefont {Clark},\ and\ \citenamefont
  {Gorshkov}}]{LR_LR2}%
  \BibitemOpen
  \bibfield  {author} {\bibinfo {author} {\bibfnamefont {Michael}\ \bibnamefont
  {Foss-Feig}}, \bibinfo {author} {\bibfnamefont {Zhe-Xuan}\ \bibnamefont
  {Gong}}, \bibinfo {author} {\bibfnamefont {Charles~W.}\ \bibnamefont
  {Clark}}, \ and\ \bibinfo {author} {\bibfnamefont {Alexey~V.}\ \bibnamefont
  {Gorshkov}},\ }\bibfield  {title} {\enquote {\bibinfo {title} {Nearly linear
  light cones in long-range interacting quantum systems},}\ }\href {\doibase
  10.1103/PhysRevLett.114.157201} {\bibfield  {journal} {\bibinfo  {journal}
  {Phys. Rev. Lett.}\ }\textbf {\bibinfo {volume} {114}},\ \bibinfo {pages}
  {157201} (\bibinfo {year} {2015})}\BibitemShut {NoStop}%
\bibitem [{\citenamefont {{Tran}}\ \emph {et~al.}(2018)\citenamefont {{Tran}},
  \citenamefont {{Guo}}, \citenamefont {{Su}}, \citenamefont {{Garrison}},
  \citenamefont {{Eldredge}}, \citenamefont {{Foss-Feig}}, \citenamefont
  {{Childs}},\ and\ \citenamefont {{Gorshkov}}}]{LR_LR3}%
  \BibitemOpen
  \bibfield  {author} {\bibinfo {author} {\bibfnamefont {M.~C.}\ \bibnamefont
  {{Tran}}}, \bibinfo {author} {\bibfnamefont {A.~Y.}\ \bibnamefont {{Guo}}},
  \bibinfo {author} {\bibfnamefont {Y.}~\bibnamefont {{Su}}}, \bibinfo {author}
  {\bibfnamefont {J.~R.}\ \bibnamefont {{Garrison}}}, \bibinfo {author}
  {\bibfnamefont {Z.}~\bibnamefont {{Eldredge}}}, \bibinfo {author}
  {\bibfnamefont {M.}~\bibnamefont {{Foss-Feig}}}, \bibinfo {author}
  {\bibfnamefont {A.~M.}\ \bibnamefont {{Childs}}}, \ and\ \bibinfo {author}
  {\bibfnamefont {A.~V.}\ \bibnamefont {{Gorshkov}}},\ }\bibfield  {title}
  {\enquote {\bibinfo {title} {{Locality and digital quantum simulation of
  power-law interactions}},}\ }\href@noop {} {\bibfield  {journal} {\bibinfo
  {journal} {ArXiv e-prints}\ } (\bibinfo {year} {2018})},\ \Eprint
  {http://arxiv.org/abs/1808.05225} {arXiv:1808.05225 [quant-ph]} \BibitemShut
  {NoStop}%
\bibitem [{\citenamefont {Lipkin}\ \emph {et~al.}(1965)\citenamefont {Lipkin},
  \citenamefont {Meshkov},\ and\ \citenamefont {Glick}}]{LMG}%
  \BibitemOpen
  \bibfield  {author} {\bibinfo {author} {\bibfnamefont {H.J.}\ \bibnamefont
  {Lipkin}}, \bibinfo {author} {\bibfnamefont {N.}~\bibnamefont {Meshkov}}, \
  and\ \bibinfo {author} {\bibfnamefont {A.J.}\ \bibnamefont {Glick}},\
  }\bibfield  {title} {\enquote {\bibinfo {title} {Validity of many-body
  approximation methods for a solvable model: (i). exact solutions and
  perturbation theory},}\ }\href {\doibase
  https://doi.org/10.1016/0029-5582(65)90862-X} {\bibfield  {journal} {\bibinfo
   {journal} {Nuclear Physics}\ }\textbf {\bibinfo {volume} {62}},\ \bibinfo
  {pages} {188 -- 198} (\bibinfo {year} {1965})}\BibitemShut {NoStop}%
\bibitem [{\citenamefont {S\o{}rensen}\ and\ \citenamefont
  {M\o{}lmer}(2000)}]{PhysRevA.62.022311}%
  \BibitemOpen
  \bibfield  {author} {\bibinfo {author} {\bibfnamefont {Anders}\ \bibnamefont
  {S\o{}rensen}}\ and\ \bibinfo {author} {\bibfnamefont {Klaus}\ \bibnamefont
  {M\o{}lmer}},\ }\bibfield  {title} {\enquote {\bibinfo {title} {Entanglement
  and quantum computation with ions in thermal motion},}\ }\href {\doibase
  10.1103/PhysRevA.62.022311} {\bibfield  {journal} {\bibinfo  {journal} {Phys.
  Rev. A}\ }\textbf {\bibinfo {volume} {62}},\ \bibinfo {pages} {022311}
  (\bibinfo {year} {2000})}\BibitemShut {NoStop}%
\bibitem [{\citenamefont {Monz}\ \emph {et~al.}(2011)\citenamefont {Monz},
  \citenamefont {Schindler}, \citenamefont {Barreiro}, \citenamefont {Chwalla},
  \citenamefont {Nigg}, \citenamefont {Coish}, \citenamefont {Harlander},
  \citenamefont {H\"ansel}, \citenamefont {Hennrich},\ and\ \citenamefont
  {Blatt}}]{PhysRevLett.106.130506}%
  \BibitemOpen
  \bibfield  {author} {\bibinfo {author} {\bibfnamefont {Thomas}\ \bibnamefont
  {Monz}}, \bibinfo {author} {\bibfnamefont {Philipp}\ \bibnamefont
  {Schindler}}, \bibinfo {author} {\bibfnamefont {Julio~T.}\ \bibnamefont
  {Barreiro}}, \bibinfo {author} {\bibfnamefont {Michael}\ \bibnamefont
  {Chwalla}}, \bibinfo {author} {\bibfnamefont {Daniel}\ \bibnamefont {Nigg}},
  \bibinfo {author} {\bibfnamefont {William~A.}\ \bibnamefont {Coish}},
  \bibinfo {author} {\bibfnamefont {Maximilian}\ \bibnamefont {Harlander}},
  \bibinfo {author} {\bibfnamefont {Wolfgang}\ \bibnamefont {H\"ansel}},
  \bibinfo {author} {\bibfnamefont {Markus}\ \bibnamefont {Hennrich}}, \ and\
  \bibinfo {author} {\bibfnamefont {Rainer}\ \bibnamefont {Blatt}},\ }\bibfield
   {title} {\enquote {\bibinfo {title} {14-qubit entanglement: Creation and
  coherence},}\ }\href {\doibase 10.1103/PhysRevLett.106.130506} {\bibfield
  {journal} {\bibinfo  {journal} {Phys. Rev. Lett.}\ }\textbf {\bibinfo
  {volume} {106}},\ \bibinfo {pages} {130506} (\bibinfo {year}
  {2011})}\BibitemShut {NoStop}%
\bibitem [{\citenamefont {del Campo}(2013)}]{PhysRevLett.111.100502}%
  \BibitemOpen
  \bibfield  {author} {\bibinfo {author} {\bibfnamefont {Adolfo}\ \bibnamefont
  {del Campo}},\ }\bibfield  {title} {\enquote {\bibinfo {title} {Shortcuts to
  adiabaticity by counterdiabatic driving},}\ }\href {\doibase
  10.1103/PhysRevLett.111.100502} {\bibfield  {journal} {\bibinfo  {journal}
  {Phys. Rev. Lett.}\ }\textbf {\bibinfo {volume} {111}},\ \bibinfo {pages}
  {100502} (\bibinfo {year} {2013})}\BibitemShut {NoStop}%
\bibitem [{\citenamefont {Campbell}\ \emph {et~al.}(2015)\citenamefont
  {Campbell}, \citenamefont {De~Chiara}, \citenamefont {Paternostro},
  \citenamefont {Palma},\ and\ \citenamefont {Fazio}}]{PhysRevLett.114.177206}%
  \BibitemOpen
  \bibfield  {author} {\bibinfo {author} {\bibfnamefont {Steve}\ \bibnamefont
  {Campbell}}, \bibinfo {author} {\bibfnamefont {Gabriele}\ \bibnamefont
  {De~Chiara}}, \bibinfo {author} {\bibfnamefont {Mauro}\ \bibnamefont
  {Paternostro}}, \bibinfo {author} {\bibfnamefont {G.~Massimo}\ \bibnamefont
  {Palma}}, \ and\ \bibinfo {author} {\bibfnamefont {Rosario}\ \bibnamefont
  {Fazio}},\ }\bibfield  {title} {\enquote {\bibinfo {title} {Shortcut to
  adiabaticity in the lipkin-meshkov-glick model},}\ }\href {\doibase
  10.1103/PhysRevLett.114.177206} {\bibfield  {journal} {\bibinfo  {journal}
  {Phys. Rev. Lett.}\ }\textbf {\bibinfo {volume} {114}},\ \bibinfo {pages}
  {177206} (\bibinfo {year} {2015})}\BibitemShut {NoStop}%
\bibitem [{\citenamefont {Campbell}\ and\ \citenamefont
  {Deffner}(2017)}]{PhysRevLett.118.100601}%
  \BibitemOpen
  \bibfield  {author} {\bibinfo {author} {\bibfnamefont {Steve}\ \bibnamefont
  {Campbell}}\ and\ \bibinfo {author} {\bibfnamefont {Sebastian}\ \bibnamefont
  {Deffner}},\ }\bibfield  {title} {\enquote {\bibinfo {title} {Trade-off
  between speed and cost in shortcuts to adiabaticity},}\ }\href {\doibase
  10.1103/PhysRevLett.118.100601} {\bibfield  {journal} {\bibinfo  {journal}
  {Phys. Rev. Lett.}\ }\textbf {\bibinfo {volume} {118}},\ \bibinfo {pages}
  {100601} (\bibinfo {year} {2017})}\BibitemShut {NoStop}%
\bibitem [{Note1()}]{Note1}%
  \BibitemOpen
  \bibinfo {note} {We note that the numerical optimization in our simulations
  may output local minima in the cost function, and thus the results presented
  in Fig. 2 are lower bounds on the optimal fidelities.}\BibitemShut {Stop}%
\bibitem [{\citenamefont {Jaschke}\ \emph {et~al.}(2017)\citenamefont
  {Jaschke}, \citenamefont {Maeda}, \citenamefont {Whalen}, \citenamefont
  {Wall},\ and\ \citenamefont {Carr}}]{Carr}%
  \BibitemOpen
  \bibfield  {author} {\bibinfo {author} {\bibfnamefont {Daniel}\ \bibnamefont
  {Jaschke}}, \bibinfo {author} {\bibfnamefont {Kenji}\ \bibnamefont {Maeda}},
  \bibinfo {author} {\bibfnamefont {Joseph~D}\ \bibnamefont {Whalen}}, \bibinfo
  {author} {\bibfnamefont {Michael~L}\ \bibnamefont {Wall}}, \ and\ \bibinfo
  {author} {\bibfnamefont {Lincoln~D}\ \bibnamefont {Carr}},\ }\bibfield
  {title} {\enquote {\bibinfo {title} {Critical phenomena and kibble–zurek
  scaling in the long-range quantum ising chain},}\ }\href
  {http://iopscience.iop.org/article/10.1088/1367-2630/aa65bc/meta} {\bibfield
  {journal} {\bibinfo  {journal} {New Journal of Physics}\ }\textbf {\bibinfo
  {volume} {19}},\ \bibinfo {pages} {033032} (\bibinfo {year}
  {2017})}\BibitemShut {NoStop}%
\bibitem [{\citenamefont {Fey}\ and\ \citenamefont {Schmidt}(2016)}]{schmidt}%
  \BibitemOpen
  \bibfield  {author} {\bibinfo {author} {\bibfnamefont {Sebastian}\
  \bibnamefont {Fey}}\ and\ \bibinfo {author} {\bibfnamefont {Kai~Phillip}\
  \bibnamefont {Schmidt}},\ }\bibfield  {title} {\enquote {\bibinfo {title}
  {Critical behavior of quantum magnets with long-range interactions in the
  thermodynamic limit},}\ }\href {\doibase 10.1103/PhysRevB.94.075156}
  {\bibfield  {journal} {\bibinfo  {journal} {Phys. Rev. B}\ }\textbf {\bibinfo
  {volume} {94}},\ \bibinfo {pages} {075156} (\bibinfo {year}
  {2016})}\BibitemShut {NoStop}%
\bibitem [{\citenamefont {Koffel}\ \emph {et~al.}(2012)\citenamefont {Koffel},
  \citenamefont {Lewenstein},\ and\ \citenamefont
  {Tagliacozzo}}]{PhysRevLett.109.267203}%
  \BibitemOpen
  \bibfield  {author} {\bibinfo {author} {\bibfnamefont {Thomas}\ \bibnamefont
  {Koffel}}, \bibinfo {author} {\bibfnamefont {M.}~\bibnamefont {Lewenstein}},
  \ and\ \bibinfo {author} {\bibfnamefont {Luca}\ \bibnamefont {Tagliacozzo}},\
  }\bibfield  {title} {\enquote {\bibinfo {title} {Entanglement entropy for the
  long-range ising chain in a transverse field},}\ }\href {\doibase
  10.1103/PhysRevLett.109.267203} {\bibfield  {journal} {\bibinfo  {journal}
  {Phys. Rev. Lett.}\ }\textbf {\bibinfo {volume} {109}},\ \bibinfo {pages}
  {267203} (\bibinfo {year} {2012})}\BibitemShut {NoStop}%
\bibitem [{\citenamefont {Vodola}\ \emph {et~al.}(2016)\citenamefont {Vodola},
  \citenamefont {Lepori}, \citenamefont {Ercolessi},\ and\ \citenamefont
  {Pupillo}}]{lrisingplus}%
  \BibitemOpen
  \bibfield  {author} {\bibinfo {author} {\bibfnamefont {Davide}\ \bibnamefont
  {Vodola}}, \bibinfo {author} {\bibfnamefont {Luca}\ \bibnamefont {Lepori}},
  \bibinfo {author} {\bibfnamefont {Elisa}\ \bibnamefont {Ercolessi}}, \ and\
  \bibinfo {author} {\bibfnamefont {Guido}\ \bibnamefont {Pupillo}},\
  }\bibfield  {title} {\enquote {\bibinfo {title} {Long-range ising and kitaev
  models: phases, correlations and edge modes},}\ }\href
  {http://stacks.iop.org/1367-2630/18/i=1/a=015001} {\bibfield  {journal}
  {\bibinfo  {journal} {New Journal of Physics}\ }\textbf {\bibinfo {volume}
  {18}},\ \bibinfo {pages} {015001} (\bibinfo {year} {2016})}\BibitemShut
  {NoStop}%
\bibitem [{\citenamefont {{Kokail}}\ \emph {et~al.}(2018)\citenamefont
  {{Kokail}}, \citenamefont {{Maier}}, \citenamefont {{van Bijnen}},
  \citenamefont {{Brydges}}, \citenamefont {{Joshi}}, \citenamefont
  {{Jurcevic}}, \citenamefont {{Muschik}}, \citenamefont {{Silvi}},
  \citenamefont {{Blatt}}, \citenamefont {{Roos}},\ and\ \citenamefont
  {{Zoller}}}]{zoller18}%
  \BibitemOpen
  \bibfield  {author} {\bibinfo {author} {\bibfnamefont {C.}~\bibnamefont
  {{Kokail}}}, \bibinfo {author} {\bibfnamefont {C.}~\bibnamefont {{Maier}}},
  \bibinfo {author} {\bibfnamefont {R.}~\bibnamefont {{van Bijnen}}}, \bibinfo
  {author} {\bibfnamefont {T.}~\bibnamefont {{Brydges}}}, \bibinfo {author}
  {\bibfnamefont {M.}~\bibnamefont {{Joshi}}}, \bibinfo {author} {\bibfnamefont
  {P.}~\bibnamefont {{Jurcevic}}}, \bibinfo {author} {\bibfnamefont
  {C.}~\bibnamefont {{Muschik}}}, \bibinfo {author} {\bibfnamefont
  {P.}~\bibnamefont {{Silvi}}}, \bibinfo {author} {\bibfnamefont
  {R.}~\bibnamefont {{Blatt}}}, \bibinfo {author} {\bibfnamefont
  {C.}~\bibnamefont {{Roos}}}, \ and\ \bibinfo {author} {\bibfnamefont
  {P.}~\bibnamefont {{Zoller}}},\ }\bibfield  {title} {\enquote {\bibinfo
  {title} {{Self-Verifying Variational Quantum Simulation of the Lattice
  Schwinger Model}},}\ }\href@noop {} {\bibfield  {journal} {\bibinfo
  {journal} {ArXiv e-prints}\ } (\bibinfo {year} {2018})},\ \Eprint
  {http://arxiv.org/abs/1810.03421} {arXiv:1810.03421 [quant-ph]} \BibitemShut
  {NoStop}%
\bibitem [{\citenamefont {Bapat}\ and\ \citenamefont {Jordan}(2018)}]{bapat}%
  \BibitemOpen
  \bibfield  {author} {\bibinfo {author} {\bibfnamefont {A.}~\bibnamefont
  {Bapat}}\ and\ \bibinfo {author} {\bibfnamefont {S.}~\bibnamefont {Jordan}},\
  }\bibfield  {title} {\enquote {\bibinfo {title} {{Bang-bang control as a
  design principle for classical and quantum optimization algorithms}},}\
  }\href@noop {} {\bibfield  {journal} {\bibinfo  {journal} {ArXiv e-prints}\ }
  (\bibinfo {year} {2018})},\ \Eprint {http://arxiv.org/abs/1812.02746}
  {arXiv:1812.02746 [quant-ph]} \BibitemShut {NoStop}%
\end{thebibliography}%

\appendix

\newpage

\onecolumngrid

\section{LMG Cost Function for $p=1$}

We evaluate
\beq
\langle + | e^{i\gamma H_I} e^{i\beta H_X} H_{LMG} e^{-i\beta H_X} e^{-i\gamma H_I}|+\rangle,
\eeq
where 
\beq
H_{LMG} = -\frac{2}{N} S_z^2 - 2g S_x
\eeq

The second piece gives 
\beq
&&-g \langle + | e^{i\gamma H_I} \sum_i X_i e^{-i\gamma H_I}|+\rangle  \\
&=&-g\langle + | \prod_{j \neq i} (\cos(\gamma)-i\sin(\gamma) Z_i Z_j) \sum_i X_i \prod_{j\neq i} (\cos(\gamma)+i\sin(\gamma) Z_i Z_j)|+\rangle
\eeq

Because any operator aside from identity and $X$ has zero expectation value in $|+\rangle$, we get contributions only from $\cos^2(\gamma)-\sin^2(\gamma)$ for each $j$.  In total, this piece is $(-gN)(\cos(2\gamma))^{N-1}$.

The first piece is
\beq
&&-\frac{N-1}{2} \langle + | e^{i\gamma H_I} e^{i\beta H_X} Z_i Z_j e^{-i\beta H_X} e^{-i\gamma H_I}|+\rangle-\frac{1}{2} \\
&=&-\frac{N-1}{2} \langle + | e^{i\gamma H_I} (\cos(2\beta)Z_i-\sin(2\beta)Y_i) (\cos(2\beta)Z_j-\sin(2\beta)Y_j) e^{-i\gamma H_I}|+\rangle-\frac{1}{2}
\eeq 

Again, we need only consider when the identity and $X$ operators arise.  One contribution to the matrix element comes from the evolution of $Y_i Z_j + Z_i Y_j$, which gives
\beq
-\sin(4\beta)\langle  + |\prod_{k\neq j,i} (\cos(\gamma)-i\sin(\gamma) Z_i Z_k)(\cos(\gamma)-i\sin(\gamma) Z_i Z_j) \\
(Y_i Z_j ) (\cos(\gamma)+i\sin(\gamma) Z_i Z_j)\prod_{k\neq j,i} (\cos(\gamma)+i\sin(\gamma) Z_i Z_k)|+\rangle \\
=\sin(4\beta) \sin(2\gamma)\cos(2\gamma)^{N-2}
\eeq

Another contribution comes from 
\beq
\sin(2\beta)^2 \langle +|\prod_{k\neq i,j}(\cos(\gamma)-i\sin(\gamma) Z_i Z_k)\prod_{l\neq i,j}(\cos(\gamma)-i\sin(\gamma) Z_j Z_l)Y_i Y_j \\
\prod_{k\neq i,j}(\cos(\gamma)+i\sin(\gamma) Z_i Z_k)\prod_{l\neq i,j}(\cos(\gamma)+i\sin(\gamma) Z_j Z_l) |+\rangle
\eeq

The transformation into two $X$ operators requires an odd number of applications of $Z_i Z_k$ and $Z_j Z_k$; each application comes with a factor of $\sin(2\gamma)^2$.  The terms which do not alter the operator come with factors of $\cos(2\gamma)^2$.  Hence, to single out the odd powers, we take the combination
\beq
(1/2)((\cos(2\gamma)^2+\sin(2\gamma)^2)^{N-2}-(\cos(2\gamma)^2-\sin(2\gamma)^2)^{N-2})\\
=(1/2)(1-\cos(4\gamma)^{N-2}).
\eeq

In total, the cost function is thus
\beq
-\frac{N-1}{4} (\sin(2\beta)^2 (1-\cos(4\gamma)^{N-2}) + 2\sin(4\beta) \sin(2\gamma)\cos(2\gamma)^{N-2})-gN(\cos(2\gamma))^{N-1}-1/2
\eeq

\section{GHZ Preparation for Even $N$}

We show below that for an even number $N$ of qubits,
\beq
|GHZ\rangle =\exp(\frac{i\pi}{4} \sum_i X_i)\exp(\frac{i\pi}{8} \sum_{ij} Z_i Z_j)\exp(\frac{3i\pi}{4N} \sum_i X_i)\exp(\frac{i\pi}{4} \sum_{ij} Z_i Z_j)|+...+\rangle.
\eeq

It is sufficient to establish 
\beq
\langle \uparrow...\uparrow|\exp(\frac{i\pi}{4} \sum_i X_i)\exp(\frac{i\pi}{8} \sum_{ij} Z_i Z_j)\exp(\frac{3i\pi}{4N} \sum_i X_i)\exp(\frac{i\pi}{4} \sum_{ij} Z_i Z_j)|+...+\rangle = \frac{1}{\sqrt{2}},
\eeq
up to a phase.  (The Ising symmetry operator is conserved as $\prod X=1$, so the matrix element for $\langle \downarrow...\downarrow|$ will also be $\frac{1}{\sqrt{2}}$. The state $|\uparrow\rangle$ is such that $Z_i |\uparrow\rangle_i = + |\uparrow\rangle_i$ and so $|\uparrow\cdots \uparrow \rangle = \prod_i |\uparrow\rangle_i$.) 

We break the matrix element in half and first the evaluate the left hand side.
First,
\beq
\exp(\frac{-i\pi}{4} \sum_i X_i)|\uparrow...\uparrow\rangle = \frac{1}{\sqrt{2^N}} \big(\prod_i (1-i X_i)\big)|\uparrow...\uparrow\rangle \\
= \frac{1}{\sqrt{2^N}} \sum_s (-i)^{(N-\sum_i z_i)/2} |z\rangle,
\eeq
where $z=\{z_1,...z_N\}$ labels a spin configuration.

Applying $\exp(\frac{-i\pi}{8} \sum_{ij} Z_i Z_j)$ and neglecting overall phase then gives
\beq
\frac{1}{\sqrt{2^N}} \sum_z \exp(-i\pi/8 \sum_{ij} z_i z_j)i^{\sum _i z_i/2} |z\rangle \\
=  \frac{1}{\sqrt{2^N}} \sum_z \exp(\frac{i\pi}{16} (-z_t^2+4z_t)) |z\rangle,
\eeq
where we have defined $z_t \equiv \sum_i z_i$.

The right hand side is:

\beq
\exp(\frac{3i\pi}{4N} \sum_i X_i)\exp(\frac{i\pi}{4} \sum_{ij} Z_i Z_j)|+...+\rangle \\
=\frac{1}{\sqrt{2^N}} \exp(\frac{3i\pi}{4N} \sum_i X_i) \sum_z \exp(\frac{i\pi}{4} \sum_{ij} z_i z_j)|z\rangle \\
=\frac{1}{\sqrt{2^N}} \prod_i (c+is X_i) \sum_z \exp(\frac{i\pi}{4} \sum_{ij} z_i z_j)|z\rangle
\eeq
where $c\equiv \cos(3\pi/4N), s\equiv \sin(3\pi/4N)$.

Consider the contributions to the coefficient of a given spin configuration $|z\rangle$.  Each contribution involves partitioning the $N$ spins into two sets $A$ and $B$ of sizes $a$ and $N-a$ respectively, and flipping the spins in set $A$.  The resulting coefficient from this given flip is 
\beq
c^{N-a} (is)^a \exp(\frac{i\pi}{4} \sum_{ij} z_i z_j) \exp(\frac{i\pi}{4} \sum_{i\in A,j\in B} (\bar{z_i}-z_i) z_j),
\eeq
where $\bar{z_i}\equiv -z_i$.

We now show that this factor only depends on the parity of $a$ (and the particular configuration $z$) and once this is fixed, the factor is independent of the partition.  The final phase factor above can be written as
\beq
\exp(\frac{-i\pi}{2} z_A (z_t -z_A)),
\eeq 
where $z_A \equiv \sum_{i\in A} z_i$.  Because $N$ is even, $z_t$ is even.  If $a$ is even, the $z_a$ is even and thus the phase factor is 1.  Moreover, it is straightforward to check that either changing the partition (keeping partition size fixed) or changing the partition size by 2 does not change the above phase.  Hence, the case of $a$ odd can be reduced to choosing $A$ to be the first spin.  The wavefunction becomes
\beq
\frac{1}{\sqrt{2^N}} \sum_z \exp(\frac{i\pi}{4} \sum_{ij} z_i z_j) \Big(\sum_{\text{even }a} {N \choose a} c^{N-a} (is)^a + \sum_{\text{odd }a} {N \choose a} c^{N-a} (is)^a \exp(\frac{-i\pi}{2} z_1 (z_t -z_1)) \Big) |z\rangle \nonumber \\ 
=\frac{1}{\sqrt{2^N}} \sum_z \exp(\frac{i\pi}{4} \sum_{ij} z_i z_j) \Big(\cos(3\pi/4)+i\sin(3\pi/4) \exp(\frac{-i\pi}{2} z_1 (z_t -z_1))  \Big) |z\rangle \nonumber
\eeq

Dropping overall phases, we get
\beq
\frac{1}{\sqrt{2^{N+1}}} \sum_z \exp(\frac{i\pi}{8} z_t^2) \Big(1-i \exp(\frac{-i\pi}{2} z_1 (z_t -z_1))  \Big) |z\rangle 
\eeq

The matrix element between left and right hand sides is thus
\beq
\frac{1}{2^N\sqrt{2}} \sum_z \exp(\frac{i\pi}{16} (3z_t^2-4z_t))(1-i \exp(\frac{-i\pi}{2} z_1 (z_t -z_1))
\eeq

Due to the last piece, any configuration with $z_t \equiv 2 (\text{mod } 4)$ does not contribute and the matrix element reduces to
\beq
\frac{1}{2^N\sqrt{2}} \sum_{z|z_t\equiv 0(\text{mod }4)} 2 \exp(\frac{i\pi}{16} (3z_t^2-4z_t)) 
=\frac{1}{\sqrt{2}}.
\eeq



\end{document}